\documentclass[acmlarge,anonymous=False]{acmart}
\usepackage{algorithm}
\usepackage{algpseudocode} 
\usepackage{multirow}
\usepackage[normalem]{ulem}
\usepackage{subcaption} 
\usepackage[table]{xcolor}
\usepackage{booktabs}
\usepackage[most]{tcolorbox}
\usepackage{caption}
\usepackage{subcaption}
\usepackage{makecell}
\usepackage{needspace}
\tcbset{
  mypromptbox/.style={
    breakable,
    colframe=black,
    colback=white,
    coltitle=white,
    colbacktitle=black,
    boxrule=0.6pt,
    arc=2pt,
    left=6pt,right=6pt,top=6pt,bottom=6pt,
    fonttitle=\bfseries,
    before upper={\raggedright},  % 添加左对齐设置
    after upper={\par}           % 确保段落正确结束
  }
}
\newtcolorbox{casebox}[2][]{%
  enhanced,
  breakable,
  sharp corners,
  boxrule=0.4pt,
  colback=gray!3!white,
  colframe=gray!60!black,
  colbacktitle=gray!6!white,
  coltitle=black,
  title={#2},
  fonttitle=\bfseries,
  borderline west={2pt}{0pt}{blue!35!black},
  opacityback=0.96,
  opacitybacktitle=0.92,
  left=8pt,right=8pt,top=6pt,bottom=6pt,
  #1
}

\AtBeginDocument{%
  }

\setcopyright{acmlicensed}
\copyrightyear{2018}
\acmYear{2018}
\acmDOI{XXXXXXX.XXXXXXX}
\acmConference[Conference acronym 'XX]{Make sure to enter the correct
  conference title from your rights confirmation email}{June 03--05,
  2018}{Woodstock, NY}
\acmISBN{978-1-4503-XXXX-X/2018/06}

\begin{document}

%%
%% The "title" command has an optional parameter,
%% allowing the author to define a "short title" to be used in page headers.
% \title{The Name of the Title Is Hope}

%%
%% The "author" command and its associated commands are used to define
%% the authors and their affiliations.
%% Of note is the shared affiliation of the first two authors, and the
%% "authornote" and "authornotemark" commands
%% used to denote shared contribution to the research.

\title{Teach Multimodal Recommendation Model to See via Personalized Visual Extraction and Adaptive Learning}

%1---------
\author{Yutong Li}
\email{lyt3612671@163.com}
% \orcid{0009-0008-6496-8850}
\affiliation{%
  \institution{Fudan University}
  \city{Shanghai}
  \country{China}
}

%2---------
\author{Xinyi Zhang}
% \orcid{0009-0008-1950-3034}
\affiliation{%
  \institution{Imperial College London}
  \city{London}
  \country{UK}}
\email{zxyzxy090588@163.com}

%3---------
\author{Ziyi Ye}
% \authornote{}
\email{zyye@fudan.edu.cn}
\affiliation{%
  \institution{Fudan University}
  \city{Shanghai}
  \country{China,}
}

\author{Daoguo Dong}
% \authornote{}
\email{dgdong@fudan.edu.cn}
\affiliation{%
  \institution{Fudan University}
  \city{Shanghai}
  \country{China,}
}

%4---------
\author{Yu-Gang Jiang}
\email{ygj@fudan.edu.cn}
\affiliation{%
  \institution{Fudan University}
  \city{Shanghai}
  \country{China}
}

%%
%% By default, the full list of authors will be used in the page
%% headers. Often, this list is too long, and will overlap
%% other information printed in the page headers. This command allows
%% the author to define a more concise list
%% of authors' names for this purpose.
\renewcommand{\shortauthors}{Yutong Li et al.}

\received{20 February 2007}
\received[revised]{12 March 2009}
\received[accepted]{5 June 2009}

%%
%% This command processes the author and affiliation and title
%% information and builds the first part of the formatted document.
% \maketitle

\begin{abstract}

Multimodal sequential recommendation (MSR) incorporates textual and visual information to improve recommendation quality. However, recent studies and our empirical analysis show that visual features are often underutilized, thereby contributing far less than textual signals. 
We attribute this issue to two factors: \textbf{insufficient visual representation learning}~(pretrained encoders fail to capture preference-relevant cues) and \textbf{unbalanced visual-text optimization}~(textual features dominate the learning process).
To address these issues, we propose Teach Multimodal \textbf{R}ecommendation Model to S\textbf{E}e via Personalized \textbf{V}isual \textbf{E}xtraction and \textbf{A}daptive \textbf{L}earning (\textbf{REVEAL}), a plug-and-play framework that enhances visual representation learning and cross-modal optimization without modifying the original recommendation backbone. 
REVEAL consists of Feedback-Guided Visual Extraction (FVE), which refines prompt-guided visual extraction through task-level feedback, and Adaptive Visual Learning (AVL), which dynamically reweights visual learning to alleviate modality imbalance. 
Experiments on multiple real-world datasets and MSR backbones demonstrate that REVEAL consistently improves recommendation performance. 
Further analysis shows that these gains arise from more effective attention to preference-relevant visual regions and better visual utilization during training.
% The code is available in the supplementary materials.
The code is available at https://github.com/YutongLi2024/REVEAL.
\end{abstract}

% insufficient extraction of preference-related visual features by pretrained encoders, and imbalanced optimization of visual representation learning across modalities during training.
% pretrained visual encoders fail to capture preference-relevant visual cues, and modality imbalance during training suppresses effective visual learning. 


% \ccsdesc[300]{Do Not Use This Code~Generate the Correct Terms for Your Paper}
% \ccsdesc{Do Not Use This Code~Generate the Correct Terms for Your Paper}
% \ccsdesc[100]{Do Not Use This Code~Generate the Correct Terms for Your Paper}

%%
%% Keywords. The author(s) should pick words that accurately describe
%% the work being presented. Separate the keywords with commas.

\maketitle
\section{Introduction}

Sequential recommendation (SR) models aim to predict future interactions by modeling the temporal dynamics of user behavior sequences~\cite{SASRec, STOSA, ChenliangLi2TOIS, YaoLina1}. Traditional SR models relying primarily on ID-based embeddings often suffer from data sparsity, which limits their capacity for deep semantic understanding and preference modeling~\cite{SR_Survey1, Zhaochunren2TOIS, XiangWangTOIS1}. To address this, multimodal sequential recommendation (MSR) integrates rich side information, such as textual and visual content, into sequential modeling. This integration enables more expressive item representations and more robust user preference capturing~\cite{MMMLP, MMSR}. Currently, most MSR models leverage pretrained textual or visual encoders to extract multimodal features~\cite{VBPR, MMSBR, VLCLIP, KDD25VLMRec}.

In MSR models, the textual modality typically conveys high-level descriptive information, whereas the visual modality provides fine-grained appearance cues. Both modalities offer indispensable insights that are difficult to derive solely from ID-based signals~\cite{MRSurvey, MRSurvey1}. Consequently, integrating both modalities is widely considered essential for robust recommendation. However, recent studies have begun to scrutinize the actual contribution of visual modality. Notably, extensive ablation studies by \citet{Limitation2Ref1} and \citet{Limitation2Ref2} consistently demonstrate that discarding visual embeddings often leads to only marginal performance degradation. Our empirical analysis, as illustrated in Figure~\ref{fig:IntroductionV1} and detailed in Section~\ref{Preliminary}, further corroborates this counterintuitive observation. Such findings suggest that visual modality remains underutilized in existing MSR frameworks, playing a surprisingly limited role in final predictions. This discrepancy raises a critical question: Why do visual modalities struggle to yield effective gains in MSR models? We attribute this bottleneck to two primary limitations:

\begin{figure*}
    \centering
    \includegraphics[width=1\linewidth]{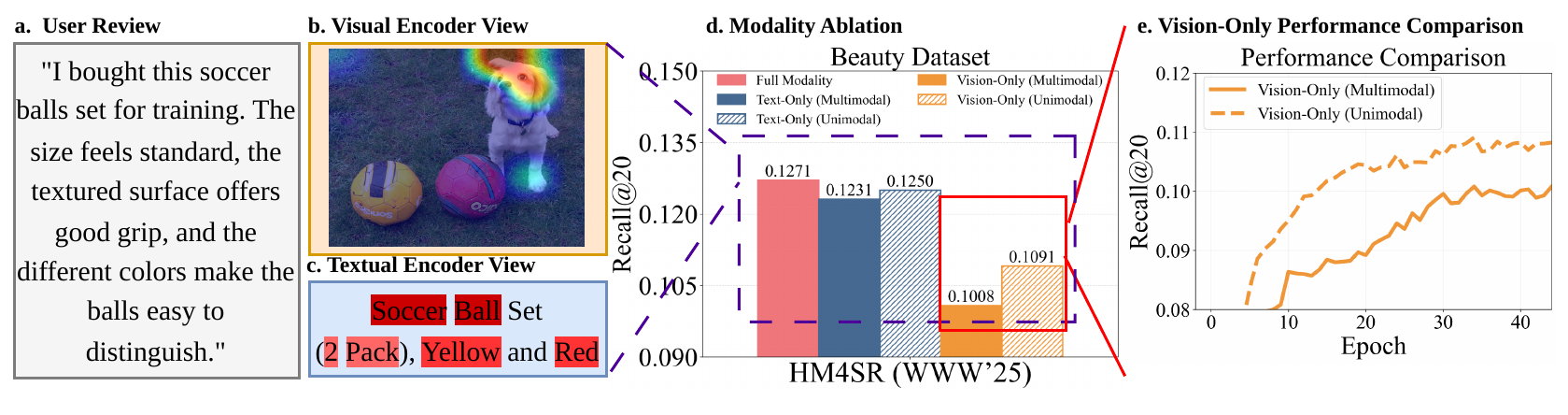}
    \caption{
    Illustration of the limitations of visual features in MSR based on the Beauty dataset.
    (a) User review of the item.
    (b) Visual encoder attention on regions weakly aligned with user preferences.
    (c) Textual encoder attention on semantics aligned with user preferences.
    (d) Modality ablation results.
    (e) Performance comparison of Vision-Only (Multimodal) vs. (Unimodal).}
    \Description{Illustration of the limited effectiveness of visual features in MSR.}
    \label{fig:IntroductionV1}
\end{figure*}

\textbf{Limitation 1: Pretrained visual encoders often fail to extract task-specific features aligned with user preferences.} Most existing MSR models employ pretrained visual encoders to derive visual representations~\cite{CLIP, Qwen2.5VL, textis}. 
Although these encoders excel at extracting general visual semantics, they often struggle to capture visual attributes highly correlated with user preferences. 
Visual modalities typically contain substantial content (e.g., complex backgrounds or decorative elements) that is irrelevant to user preference~\cite{TextAndImage2_BeFA, MDSBR}. 
Since these encoders are trained on large-scale generic datasets to learn broad visual patterns, they tend to prioritize visually dominant regions while overlooking subtle attributes that drive user decision-making~\cite{CLIP, MDSBR, textis}. 
In contrast, textual modalities with key item attributes (e.g., titles and categories) enable textual encoders to readily capture preference-relevant semantics~\cite{Balance1, Balance2}. 
As illustrated in Figure ~\ref{fig:IntroductionV1}(a-c), the visual encoder is often distracted by irrelevant regions that occupy large image areas (e.g., a background pet), whereas the textual encoder focuses on attributes consistent with user reviews. Consequently, when MSR models employ frozen or weakly adapted visual encoders, the resulting embeddings provide limited discriminative power for the recommendation task.

\textbf{Limitation 2: Existing MSR models suffer from modality imbalance, impeding the effective learning of visual information.} 
As illustrated in Figure~\ref{fig:IntroductionV1}(d)(e) and detailed in Section~\ref{Preliminary}, the textual modality is more readily exploited by MSR models and tends to dominate the training process, which hampers the model's ability to adequately learn and utilize visual signals~\cite{Balance1, Balance2}. The underlying cause of this imbalance lies in the nature of information carried by each modality. While textual data directly encodes attributes highly relevant to user decision-making (e.g., item titles and categories), visual data contains more dispersed information and higher levels of noise. Under these conditions, the training process is often dominated by the more "informative" textual modality. This dominance leads the model to rely primarily on textual signals, consequently suppressing the optimization and sufficient learning of the visual modality~\cite{New_Balance1, New_Balance2, New_Balance3}. 

To address these limitations, we propose Teach Multimodal \textbf{R}ecommendation Model to S\textbf{E}e via Personalized \textbf{V}isual \textbf{E}xtraction and \textbf{A}daptive \textbf{L}earning (REVEAL), a \textbf{plug-and-play} module comprising two components designed to improve visual feature extraction and optimization, respectively. First, \textbf{Feedback-Guided Visual Extraction (FVE)} enhances the task-relevance of visual representations through user-guided feedback. Inspired by the principles of gradient propagation~\cite{gradient_propagation}, FVE leverages task-level feedback from recommendation outcomes to iteratively refine the textual prompts fed into VLM. By rewriting the initial prompt into a more preference-aware form, FVE steers the VLM to focus on attributes prioritized by users, thereby extracting preference-relevant features without requiring costly fine-tuning. Second, \textbf{Adaptive Visual Learning (AVL)} employs a gradient reweighting mechanism to alleviate the under-optimization of the visual modality. By dynamically adjusting the contribution of visual signals during backpropagation, AVL intensifies the involvement of visual features in model updates, mitigating modality imbalance and preventing visual information from being overshadowed by dominant textual signals. We evaluate REVEAL in the context of SR, a paradigm widely adopted in industrial applications~\cite{DIN, DIEN}. Extensive experiments demonstrate that our framework integrates seamlessly with various MSR backbones across multiple datasets, yielding consistent and significant performance gains.

Our main contributions are summarized as follows:
\begin{itemize}
\item We conduct a systematic analysis of why the visual modality remains under-utilized in existing MSR models. Our analysis uncovers two critical limitations: the gap in preference-relevant feature extraction and the presence of modality imbalance during training.

\item We propose REVEAL, a plug-and-play framework designed to revitalize visual modality. It comprises two key components: FVE, which aligns visual representations with user preferences via task-level feedback, and AVL, which employs gradient reweighting to explicitly mitigate the under-optimization of visual signals.

\item Through rigorous experiments on multiple real-world datasets and diverse MSR backbones, we demonstrate that REVEAL consistently improves performance and exhibits robust compatibility.
\end{itemize}

\section{Preliminary}\label{Preliminary}

We conduct a preliminary ablation study on the Amazon-Beauty dataset based on the HM4SR model to empirically validate the observations presented in the Introduction. Following standard protocols for multimodal recommendation, we ensure a fair and comparable evaluation by adhering to established experimental settings~\cite{Limitation2Ref1, Limitation2Ref2}. For feature extraction, we adopt the representative pre-trained vision-language model CLIP-ViT-B-32\footnote{\url{https://huggingface.co/sentence-transformers/clip-ViT-B-32}}
 to obtain both textual and visual features.

Figure~\ref{fig:IntroductionV1}(d) illustrates the performance across five input configurations. For unimodal models, absent modalities are replaced with zero vectors to eliminate information from the missing modality while maintaining input dimensionality. This design allows us to fairly evaluate the actual contribution of each modality under missing-modality conditions without changing the model architecture or parameter scale, and is also consistent with prior work~\cite{Limitation2Ref1, Limitation2Ref2}. For multimodal models trained on both textual and visual features, we report both full-modality performance and unimodal performance (obtained by masking one modality during inference), consistent with prior research~\cite{Limitation2Ref1, Limitation2Ref2, Balance2}.

\begin{center}
\begin{tcolorbox}[
    width=0.96\linewidth,
    colback=gray!3!white,
    colframe=gray!60!black,
    % colbacktitle=gray!6!white,
    boxrule=0.5pt,
    arc=2mm,
    left=8pt,
    right=8pt,
    top=6pt,
    bottom=6pt,
    title=\textbf{Key Definition of Experiment Settings},
    fonttitle=\bfseries
]
\textbf{Unimodal:} The model is trained without access to the other modality at all.

\textbf{Multimodal:} The model is trained with both modalities, but one modality is masked only at inference time.
\end{tcolorbox}
\end{center}

The configurations are summarized as follows: (1) \textbf{Full Modality}: A standard multimodal setting utilizing both textual and visual modalities for training and inference. (2) \textbf{Text-Only (Unimodal)}: Utilizing only the textual modality for both training and inference. (3) \textbf{Text-Only (Multimodal)}: Training with both modalities but replacing visual inputs with zero vectors during inference. (4) \textbf{Vision-Only (Unimodal)}: Utilizing only the visual modality for both training and inference. (5) \textbf{Vision-Only (Multimodal)}: Training with both modalities but replacing textual inputs with zero vectors during inference.

There are three phenomena highlighting the limited effectiveness of the visual modality in MSR: (1) In the unimodal setting, models utilizing only visual modality significantly underperform those using only textual modality. This suggests that within current MSR frameworks, visual features are inherently weaker and less discriminative than textual features. (2) Comparing Full Modality, Text-only (Multimodal), and Vision-only (Multimodal) reveals that removing visual inputs during inference leads to only a marginal performance drop. In contrast, removing textual inputs results in a substantial decline, suggesting that visual features contribute limited gains to the final recommendation. (3) The performance gap between the multimodal and unimodal settings is significantly larger for the visual modality than for the textual modality (i.e., Vision-only (Multimodal) vs. (Unimodal) compared to Text-only (Multimodal) vs. (Unimodal)). This indicates that the textual modality dominates the training process, leaving visual modality severely under-optimized. These phenomena are consistently observed across various MSR architectures, datasets, and evaluation metrics, rather than being confined to a specific model or benchmark.
\section{Related Work}

\subsection{Sequential recommendation}

Sequential recommendation (SR) aims to predict items that a user is likely to interact with in the future based on their historical interaction sequence. Early studies primarily modeled user behavior dynamics as Markov processes, capturing short-term dependencies through probabilistic formulations~\cite{FPMC, he2016fusing, Zhaochunren1WWW26}. With the advent of deep learning, researchers began to leverage recurrent neural networks (RNNs) and convolutional neural networks (CNNs) to model user behavior sequences, achieving substantial performance improvements. For example, GRU4Rec was the first to introduce gated recurrent units into SR~\cite{GRU4Rec}, while NARM employed a dual-GRU architecture to separately capture sequential patterns and users’ main interests~\cite{NARM}. However, these approaches still suffer from limitations in modeling long-range dependencies. In recent years, Transformer-based architectures with self-attention mechanisms have demonstrated strong effectiveness in SR. SASRec pioneered the adoption of self-attention to model long-term dependencies within user interaction sequences~\cite{SASRec}. Subsequently, BERT4Rec employed a bidirectional self-attention encoder to capture richer contextual information~\cite{Bert4rec}. Building upon these advances, a series of extensions have been proposed. For instance, SSE-PT addresses the lack of personalization in Transformers by introducing randomly shared embeddings~\cite{SSE-PT}. Currently, SR methods based on self-attention and graph neural networks have become dominant paradigms in this field, continuously pushing the performance frontier~\cite{STOSA, S3Rec, SURGE}. Despite this progress, most existing methods rely solely on item ID sequences and fail to fully exploit multimodal content information. Motivated by this limitation, we focus on MSR to better capture user interests through multimodal signals.

\subsection{Multimodal Sequential Recommendation}

Traditional sequential recommendation methods primarily model user--item interaction sequences, making it difficult to fully exploit the rich multimodal content of items, such as textual descriptions and visual information. Multimodal sequential recommendation (MSR) addresses this limitation by integrating multimodal content features to alleviate data sparsity in recommendation systems, thereby improving both performance and model robustness~\cite{ChenliangLi1TOIS, MMSBR}. A growing body of work has explored different strategies for incorporating multimodal information. UniSRec leverages textual signals to learn transferable item representations across scenarios, reducing reliance on explicit user IDs~\cite{UniSRec}. More recently, Transformer-based MSR methods have achieved notable progress. MMSR integrates homogeneous and heterogeneous multimodal user--item interaction features through graph modeling and dual-attention mechanisms~\cite{MMSR}, while MMMLP demonstrates the effectiveness of a lightweight multilayer perceptron architecture for large-scale MSR~\cite{MMMLP}. In addition, several studies have introduced pretraining and prompt-learning techniques. For example, MISSRec adopts a multimodal pretraining and transfer learning framework to dynamically generate user-adaptive item representations~\cite{MISSRec}, whereas HM4SR employs a two-layer mixture-of-experts model to extract user interest-relevant information and leverages explicit temporal signals to capture dynamic interest evolution~\cite{HM4SR}. Despite these advances in multimodal fusion, existing methods still face two major challenges. First, visual feature extraction is often decoupled from user preferences and lacks adaptive adjustment to personalized needs. Second, during joint multimodal training, optimization imbalance across modalities hinders the effective learning of visual representations. To address these issues, We propose REVEAL to effectively utilize visual features.

\section{Methodology}

\begin{figure*}
    \centering
    \includegraphics[width=1\linewidth]{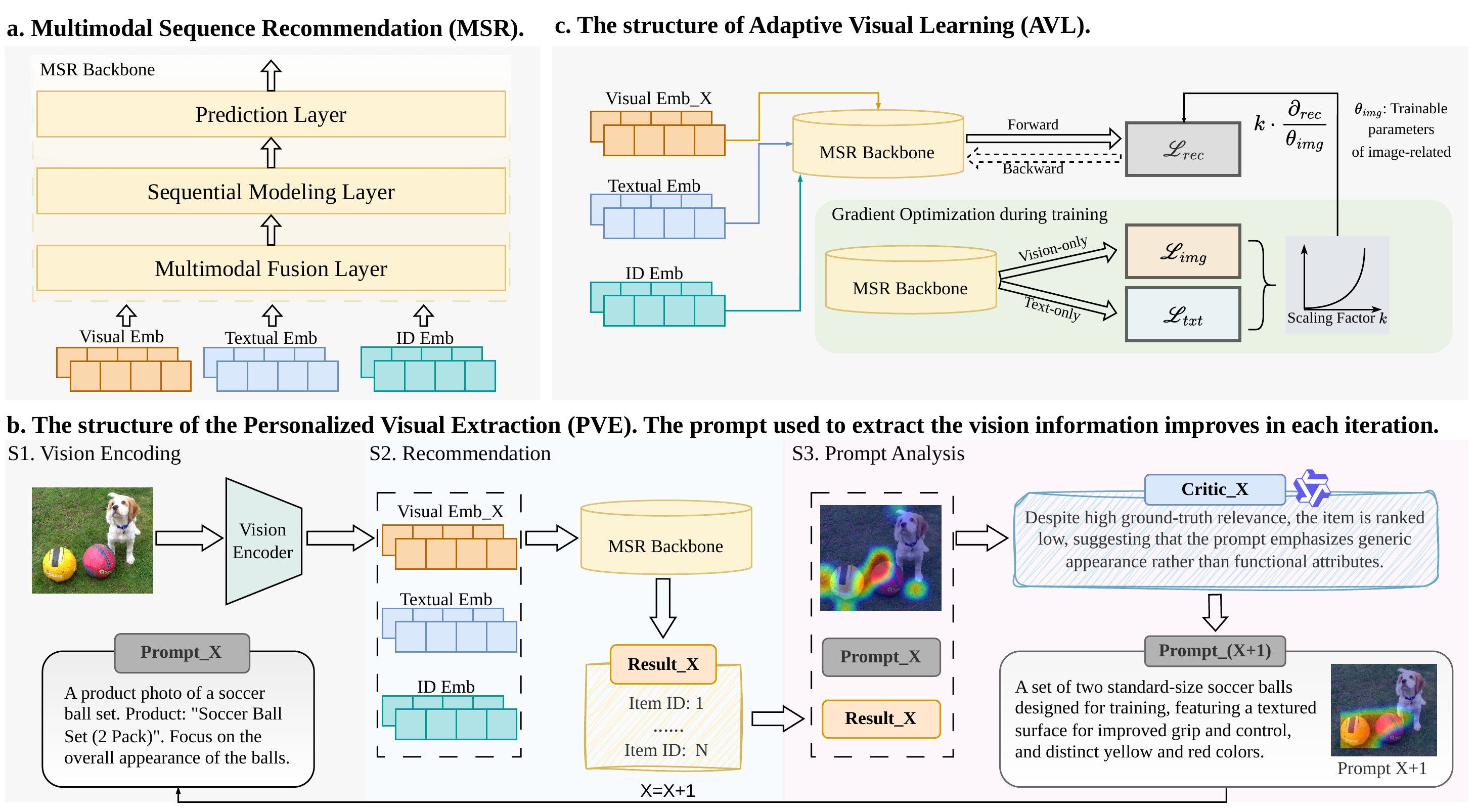}
    \caption{The overall architecture of the proposed REVEAL.}
    \Description{The method figure.}
    \label{fig:method}
\end{figure*}

\subsection{Problem Definition}
MSR aims to exploit multimodal information and users’ historical behaviors to generate personalized recommendations for their next interactions. 
Let $\mathcal{U}$ denotes the user set and $\mathcal{P}$ the item set. 
User-item interactions are temporally ordered to form behavior sequences. 
Specifically, for each user $u \in \mathcal{U}$, its interaction history is represented as an ordered sequence $\mathcal{S}^u = [p^u_1, p^u_2, \ldots, p^u_{|\mathcal{S}^u|}]$, where $p^u_i \in \mathcal{P}$ corresponds to the item interacted at the $i$-th interaction step. 
Consistent with prior MSR studies~\cite{MISSRec, M3SRec, HM4SR}, we consider visual and textual modalities as the primary sources of side information. Formally, each item $p_i$ is associated with a triplet representation
$p_i = \{p_i^{id}, p_i^{v}, p_i^{t}\}$,
comprising its identifier, visual content, and textual description. 
The objective of MSR is to jointly capture the temporal dynamics contained in user interaction sequences and the multimodal information of items, thereby accurately predicting the next item a user will interact with.

% 2. Overview FraneWork 
% 围绕图片讲一下整体流程

% 本节首先介绍标准的多模态序列推荐（MSR）流程，以说明 TRSRec 如何在不改变主干结构的前提下无缝集成到现有模型中。然后我们概述 TRSRec 的整体设计和详细阐述PRISM的技术细节。
% 如图~\ref{fig:method}(a)所示，用户的交互序列首先由ID嵌入层处理，以捕获协作信号，生成序列的$\mathbf{e}^{id}$~\cite{SASRec,Bert4rec,STOSA}。Item的多模态信息采用模态特定的编码器来处理 $p^{img}$ 和 $p^{txt}$，并计算相应的潜在嵌入，如 $\mathbf{e}^{img} = \operatorname{imgEmb}(p^{img}), \mathbf{e}^{txt} = \operatorname{textEmb}(p^{txt})$，其中 $\operatorname{textEmb}(\cdot)$ 与原MSR Backbone使用模型一致. $\operatorname{imgEmb}(\cdot)$ 选择为广泛使用的VLM(e.g., Qwen2.5-VL ), 然后，为每个项目分配一个位置嵌入，以指示其在序列中的相对或绝对位置。然后送入到Multimodal Fusion Layer 来融合图像（img）和文本（txt）。由此得到的融合项目表示（将ID嵌入与多模态信号相结合）随后被输入到序列编码器中，该编码器遵循SASRec和STOSA等模型的架构，以捕获序列依赖关系和用户兴趣。

% TRSRec通过优化特征提取和MSR的训练流程来帮助MSR模型更好地利用图像特征. 具体来说,我们的框架分为FVE和AVL两个部分. 为了解决第一个Limitation, 我们提出FVE模块(如图2b所示), 我们提出FVE模块,它作用与Embedding Layer，在训练阶段利用用户反馈引导视觉表示学习。FVE 通过可优化的 Prompt 控制 VLM 的图像关注方式，并以预测排序与真实交互之间的偏差作为监督信号，对 Prompt 进行迭代更新，从而促使 VLM 提取更具偏好相关性的视觉特征。该过程不涉及视觉编码器参数更新，且不引入额外的推理开销。为了解决第二个 Limitation，我们提出了 AVL 模块(如图 2c 所示),它作用于MSR Backbone, 在训练阶段基于单模态排名损失刻画模态间的相对优化状态，并据此对推荐损失的反向传播梯度进行自适应重加权，以缓解模态竞争带来的梯度分配失衡，增强图像模态的学习效果。该过程不引入额外监督，模型始终由推荐损失统一优化。

\subsection{Framework Overview}
This section briefly reviews the standard MSR pipeline to clearly illustrate how the proposed REVEAL can be seamlessly integrated into existing MSR models without altering their backbone architectures. 
We then introduce its core components in detail. 

As shown in Figure~\ref{fig:method}(a), a user--item interaction sequence is first processed by an ID embedding layer to capture collaborative signals, producing the sequential representation $e^{id}$ for each item~\cite{SASRec,Bert4rec,STOSA}. 
The multimodal information of each item is encoded by modality-specific encoders for visual and textual inputs $p^{v}$ and $p^{t}$, yielding latent representations $e^{v}$ and $e^{t}$. 
The textual encoder is identical to that used in the original MSR backbone, while the visual encoder is instantiated by widely used VLMs~(e.g., Qwen2.5-VL). 
Positional embeddings are added to encode each item’s relative or absolute position in the sequence. 
The generated multimodal embeddings are fused and fed into a sequential modeling layer following the architectural design of SASRec~\cite{SASRec} or STOSA~\cite{STOSA} to model sequential dependencies and user interest dynamics.

However, existing models simply vectorize and fuse different modalities and neglect the limitations of visual modality. REVEAL utilizes the \textbf{Feedback-Guided Visual Extraction (FVE)} component to tackle the first limitation and \textbf{Adaptive Visual Learning (AVL)} component to address the second limitation.
% TRSRec enhances the utilization of visual information by jointly optimizing visual feature extraction and training dynamics. 
% The framework consists of two complementary modules, \textit{Feedback-Guided Visual Extraction (FVE)} and \textit{Adaptive Visual Learning (AVL)}. 
% To address the first limitation, 
As shown in Figure~\ref{fig:method}(b), FVE guides the extraction of visual features. 
It utilizes learnable prompts optimized through users' personalized feedback to modulate the VLM, enabling the extraction of preference-relevant visual features. 
The optimization of the learnable prompts does not update the visual encoder and introduces no additional inference overhead. 
As shown in Figure~\ref{fig:method}(c), AVL is applied to the training process of MSR backbone. It estimates the relative optimization states of different modalities using unimodal recommendation losses and adaptively reweights gradients from the recommendation loss. AVL strengthens visual learning while preserving a unified optimization objective to alleviate modality imbalance. This process introduces no auxiliary supervision and optimizes the learning of MSR solely with gradient-based operation.

% 3. Feedback-Guided Visual Extraction
% Feedback-Guided Visual Extraction (FVE) 旨在通过训练阶段对视觉特征提取提示语的迭代优化，使图像表征逐步对齐用户偏好。在整个训练过程中，视觉语言模型的参数保持不变，仅对提示进行迭代优化。 如图 2(b) 所示，FVE的优化流程由Image Encoding、Recomendation和Prompt Analysis三个阶段组成。

% 在第 X 次迭代中，我们使用固定的视觉模型 (VLM) 执行提示引导的图像编码。具体来说，给定产品图像 $\mathbf{p}^{img}$ 和当前提示 $\text{提示 X}$，视觉编码器提取的图像表示如下：

% \begin{equation}
% \mathbf{e}^{img\,X} = \text{VLM}\big( \mathbf{p}^{img}, \text{提示 X} \big),
% \end{equation}

% 其中，$\text{VLM}(\cdot)$ 表示预训练的 VLM，本文采用 Qwen2.5-VL-7B。提示 X 由商品级元数据（例如，标题、品牌和类别）以及与该商品相关的用户评论构建而成, 为了引导编码器提取符合用户偏好的视觉特征。用于初始化提示 X 的提示模板如图~\ref{fig:prompt_x} 所示。需要注意的是，视觉编码器的参数在训练过程中保持不变。

\subsection{Feedback-Guided Visual Extraction}\label{sec:FVE}
Feedback-Guided Visual Extraction (FVE) aims to gradually improve the preference relevance of visual representations through iterative prompt refinement guided by recommendation feedback. Rather than maintaining one identical prompt for all users and items, FVE adopts a shared prompt template schema and refinement mechanism, while the instantiated prompt content is dynamically constructed from the context of the current user-item interaction. Throughout the optimization process, the parameters of the VLM remain fixed, while only the instantiated prompt is iteratively refined.

As shown in Figure~\ref{fig:method}(b), FVE is implemented as an alternating optimization process with a coarser granularity than backbone parameter updates. Specifically, under a fixed prompt, the MSR backbone is first trained to convergence. This means that one refinement cycle corresponds to one complete backbone training stage under the current prompt, followed by one prompt analysis and refinement step. The resulting recommendation outputs are then analyzed to evaluate the quality of the current prompt and refine it for the next training cycle. Therefore, each refinement cycle consists of three stages: Vision Encoding, Recommendation, and Prompt Analysis. Since FVE does not maintain a fixed user-specific prompt for storage, its additional storage overhead remains limited.

\subsubsection{Vision Encoding and Recommendation}
At refinement cycle $x$, a frozen VLM is first employed to extract the visual embedding under the guidance of the current instantiated prompt $\text{Prompt\_X}$:
\begin{equation}
e^{v\_X} = \text{VLM}\big(p^{v},\; \text{Prompt\_X} \big).
\end{equation}

The resulting visual embedding is then jointly fed with $e^{id}$ and $e^{t}$ into the MSR backbone trained under the current prompt to obtain recommendation predictions $\text{Result\_X}$:
\begin{equation}
\text{Result\_X} = f_{\text{MSR}}\big(e^{v\_X},\; e^{id},\; e^{t}\big).
\end{equation}

\subsubsection{Prompt Analysis}
In recommendation systems, ground-truth user interactions provide explicit supervision for determining whether recommendation results are correct. Instead of relying solely on numerical loss functions, we perform a textual evaluation of the visual representation $e^{v\_X}$ by comparing the ground truth $G$ and the recommendation result $\text{Result\_X}$. Specifically, at refinement cycle $x$, the visual representation $e^{v\_X}$ is first extracted using the current instantiated prompt $\text{Prompt\_X}$, and the MSR backbone then produces the corresponding recommendation result $\text{Result\_X}$. Given the current prompt, evaluation instruction, ground-truth interactions, and recommendation outcome, we employ a VLM to jointly perform evaluation and analysis:
\begin{equation}
\text{Critic\_X} \gets \text{VLM}\big(\text{Prompt\_X},\; \text{EvalInstr},\; G,\; \text{Result\_X}\big).
\end{equation}

The evaluation instruction $\text{EvalInstr}$ explicitly defines the evaluation criteria and guides the VLM to analyze the recommendation result with respect to the ground truth. The analysis focuses on whether the visual embedding extracted under the current prompt $\text{Prompt\_X}$ captures visual signals that are useful for preference modeling in the current recommendation context. When recommendation errors occur, the critic further explains why the current prompt fails to provide effective visual guidance. In particular, $\text{Critic\_X}$ identifies deficiencies in the current prompt design, such as overemphasizing coarse-grained appearance cues or overlooking preference-relevant visual attributes. The prompt is then refined through a textual gradient-descent-style update:
\begin{equation}
\text{Prompt\_(X+1)} = \text{VLM}\big(\text{Critic\_X},\; \text{Prompt\_X},\; p^{v}\big).
\end{equation}

Through this iterative optimization, the prompt is progressively refined under recommendation feedback supervision, enabling more preference-aware and feedback-guided visual feature extraction without updating the parameters of the VLM. This optimization procedure follows the principle of gradient propagation~\cite{gradient_propagation}, where numerical gradients are replaced by natural language feedback to guide prompt refinement. The refinement process is repeated for a fixed number of cycles $X_{\max}$, where $X_{\max}$ is a tunable hyperparameter. 

The templates of the initial $\text{Prompt\_X}$ and the evaluation instruction $\text{EvalInstr}$ are presented below:

\Needspace{10\baselineskip}
\begin{casebox}{Initial $\text{Prompt\_X}$ Template}
\textbf{Input fields from \texttt{meta data} and the review set associated with the current user-item interaction context:}
\begin{itemize}
  \item \texttt{title}
  \item \texttt{categories}
  \item high-frequency \texttt{keywords} extracted from the associated review set
\end{itemize}
\textbf{$\text{Prompt\_X}$ template:}
\begin{quote}
An item image of a <\texttt{category}>. Item name: <\texttt{title}>. Users often mention: <\texttt{keywords}>. Focus on the overall appearance of the <\texttt{item}>.
\end{quote}
\end{casebox}

\begin{casebox}{Evaluation Instruction $\text{EvalInstr}$ Template}
\textbf{Input fields from recommendation and ground truth:}
\begin{itemize}
  \item Recommendation results $\text{Result\_X}$
  \item Ground truth interactions $G$
\end{itemize}

\textbf{$\text{EvalInstr}$ template:}
\begin{quote}
Assess whether the current prompt effectively extracts visual features for recommendation by comparing the prediction with the ground truth. If recommendation errors occur, analyze whether the prompt misses preference-relevant visual attributes and suggest improvements to the prompt.
\end{quote}
\end{casebox}

\subsubsection{Prompt Construction and Granularity}
FVE does not use one identical prompt for all users and items. Instead, it adopts a shared prompt template schema and refinement mechanism, while the instantiated prompt content is dynamically constructed from the metadata and review-derived signals of the current user-item interaction instance. Specifically, the initial prompt is formed using the item's meta data, including \texttt{title} and \texttt{categories}, together with top-$k$ high-frequency keywords extracted from the review set associated with the current user-item interaction instance. Therefore, the prompt is instance-conditioned rather than globally fixed, allowing visual extraction to better reflect preference-relevant signals in the current recommendation context.

To obtain the review-derived keywords, we first collect the review texts associated with the current user-item interaction instance, and then apply basic text preprocessing, including lowercasing, punctuation removal, and stop-word filtering. After that, keyword frequency is computed, and the top-$5$ most frequent terms are retained to instantiate the prompt template. In this way, the prompt content captures interaction-relevant preference cues without persistently storing a separate user-specific prompt.

Under the instantiated prompt $\text{Prompt\_X}$, the frozen VLM extracts the visual embedding of the current item image. The resulting recommendation output is then compared with the ground truth to generate textual feedback, which is further used to refine the prompt for the next refinement cycle. Therefore, the prompt refinement process in FVE should be understood as feedback-guided and instance-conditioned, rather than as a user-specific prompt storage mechanism.

\subsection{Adaptive Visual Learning}

Adaptive Visual Learning (AVL) aims to alleviate the under-opt-imization of visual features by explicitly reweighting modality-specific gradients during backpropagation. Without modifying the MSR backbone or fusion architecture, AVL estimates the relative influence of visual and textual signals on recommendation outcomes and adaptively calibrates their gradient contributions to model updates.

To approximately assess the relative impact of different modalities on the final prediction, we construct two alternative evaluations that differ only at the input level: \textbf{Vision-only:} visual embeddings are retained, while textual embeddings are replaced with zero vectors. \textbf{Text-only:} textual embeddings are retained, while visual embeddings are replaced with zero vectors. Under these settings, the corresponding recommendation results are denoted as $s^{v}$ and $s^{t}$, which approximately reflect the ability of model to leverage visual and textual modalities within the current MSR backbone:
\begin{equation}
s^{v}=f_{\text{MSR}}\big({e}^{id},\; {e}^{v},\; {0}\big),
\end{equation}
\begin{equation}
s^{t}=f_{\text{MSR}}\big({e}^{id},\; {0},\;{e}^{t}\big).
\end{equation}
${0}$ denotes a zero vector with the same dimensionality as the masked modality. This masking strategy simulates vision-only or text-only conditions while keeping the model architecture and computational pathway unchanged, thereby isolating the impact of each modality on the recommendation performance~\cite{Limitation2Ref1, Limitation2Ref2}.

\subsubsection{Unimodal Loss}

To remain consistent with the main optimization objective, we directly reuse the loss function $\mathcal{L}_{\text{MSR}}$ employed by the MSR backbone and compute the losses under the unimodal input conditions as follows:
\begin{equation}
\mathcal{L}_{v}=\mathcal{L}_{\text{MSR}}\big(s^{v}_{u,i^+},\; s^{v}_{u,i^-}\big),
\end{equation}
\begin{equation}
\mathcal{L}_{t}=\mathcal{L}_{\text{MSR}}\big(s^{t}_{u,i^+},\; s^{t}_{u,i^-}\big).
\end{equation}
where $u$ denotes the user, $i^+$ and $i^-$ represent the positive item interacted by the user and a randomly sampled negative item, respectively. $\mathcal{L}_{\text{MSR}}(\cdot)$ denotes the original recommendation loss adopted by the MSR backbone. Regardless of its specific form (e.g., BPR loss or cross entropy loss), the internal computation remains unchanged, with only the input scores replaced by the corresponding unimodal predictions.

After averaging over a batch, we obtain the unimodal loss statistics $\mathcal{L}_{v}$ and $\mathcal{L}_{t}$. A smaller loss indicates that, under the current parameters, the corresponding modality can more easily discriminate positive and negative samples, reflecting the stronger unimodal contribution.

\subsubsection{Gradient Optimization}

Based on the unimodal losses, our goal is to quantify the contribution disparity between the visual and textual modalities at the current training state and dynamically calibrate the optimization process accordingly. At each training iteration $x$, we obtain the unimodal recommendation losses $\mathcal{L}_x^{v}$ and $\mathcal{L}_x^{t}$. To quantify the loss gap between the visual and textual modalities at the current training step, we define:
\begin{equation}
\Delta_x^{v}=\mathcal{L}_x^{v} - \mathcal{L}_x^{t}.
\end{equation}
$\Delta_x^{v}>0$ indicates that recommendation based solely on visual information is more challenging than that based on textual information. 

We further map this difference $\Delta_x^{v}$ to a strictly positive scaling factor via an exponential function:
\begin{equation}
r_x^{v}=\exp\left(\Delta_x^{v}\right).
\end{equation}
The value $r_x^{v}$ captures the performance gap between the visual and textual modalities for the current batch of data, which further serves as a key signal for dynamic calibration.

Due to the sparsity of user interactions and the randomness introduced by negative sampling, modality disparities observed in a single batch can be highly volatile. Directly adjusting gradients based on $r_x^{v}$ alone may therefore lead to unstable training. To obtain a more robust reference, we maintain cumulative statistics of the unimodal losses observed throughout training:
\begin{equation}
\bar{\mathcal{L}}_x^{v}=\frac{1}{x}\sum_{\tau=1}^{x}\mathcal{L}_\tau^{v},\quad
\bar{\mathcal{L}}_x^{t}=\frac{1}{x}\sum_{\tau=1}^{x}\mathcal{L}_\tau^{t}.
\end{equation}

These statistics reflect the average optimization difficulty encountered when the MSR model relies primarily on a single modality to distinguish positive and negative samples. Accordingly, we define a reference modality gap as:
\begin{equation}
\bar{r}_x^{v}=\exp\left(\bar{\mathcal{L}}_x^{v}-\bar{\mathcal{L}}_x^{t}\right).
\end{equation}
$\bar{r}_x^{v}$ represents the average performance gap between the visual and textual modalities accumulated up to the current training stage.

Finally, AVL calibrates the gradient updates of vision-related parameters based on the deviation between the instantaneous and historical modality gaps:
\begin{equation}
K_x^{v}=\exp\left(\alpha\cdot\left(r_x^{v}-\bar{r}_x^{v}\right)\right),
\end{equation}
where $\alpha>0$ is a hyperparameter controlling the strength of calibration. The design of $K_x^{v}$ is grounded in the current batch-specific gap $r_x^{v}$ with the historical average $\bar{r}_x^{v}$. 
If the visual modality performs worse than average~(i.e., $r_x^{v}>\bar{r}_x^{v}$), then $K_x^{v}>1$, amplifying the gradients of vision-related parameters; otherwise, $K_x^{v}<1$, resulting in suppressed updates. With the calibration factor $K_x^{v}$, AVL rescales the backpropagated gradients associated with the visual modality without altering the original training objective of MSR backbone.

Let $\theta$ denote the set of model parameters, and $\theta^{v}$ the subset of parameters associated with visual modality. The gradient adjustment is defined as:
\begin{equation}
\frac{\partial \mathcal{L}_{\text{MSR}}}{\partial \theta^{v}}
\leftarrow
K_x^{v}\cdot
\frac{\partial \mathcal{L}_{\text{MSR}}}{\partial \theta^{v}}.
\end{equation}

Accordingly, the parameter update at iteration $x$ becomes:
\begin{equation}
\theta^{v}_{x+1}
=
\theta^{v}_{x}
-
\eta\,K_x^{v}\,
\frac{\partial \mathcal{L}_{\text{MSR}}}{\partial \theta^{v}_{x}},
\end{equation}
where $\eta$ denotes the learning rate.

\begin{algorithm}[t]
\caption{Feedback-Guided Visual Extraction (FVE)}
\label{alg:FVE}
\begin{algorithmic}[1]
\Require Item visual content $p^{v}$, ID embedding $e^{id}$, textual embedding $e^{t}$, ground truth $G$, frozen ${\text{VLM}}$, MSR backbone $f_{\text{MSR}}$, evaluation instruction $\text{EvalInstr}$, target iteration $X_{\max}$
\Ensure Optimized prompt $\text{Prompt\_(X+1)}$

\State Initialize $\text{Prompt\_X}$ \Comment{Extracted from item metadata and user reviews}
\State Set $X \gets 0$ \Comment{Current iteration counter}

\While{$X < X_{\max}$}
    \State \textbf{Step 1: Encoding Visual Content}
    \State $e^{v\_X} \gets {\text{VLM}}(p^{v}, \text{Prompt\_X})$
    
    \State \textbf{Step 2: Recommendation}
    \State $\text{Result\_X} \gets f_{\text{MSR}}(e^{v\_X}, {e}^{id}, {e}^{t})$

    \State \textbf{Step 3: Evaluation and Analysis}
    \State $\text{Critic\_X} \gets {\text{VLM}}\big(\text{Prompt\_X}, \text{EvalInstr}, {G}, \text{Result\_X}\big)$

    \State \textbf{Step 4: Prompt Update}
    \State $\text{Prompt\_(X+1)} \gets {\text{VLM}}(\text{Critic\_X}, \text{Prompt\_X}, p^{v})$
    
    \State $X \gets X + 1$ \Comment{Advance to next iteration}
\EndWhile

\State \Return $\text{Prompt\_(X+1)}$
\end{algorithmic}
\end{algorithm}

% 5. 训练和优化。
% \subsection{Training and Inference}
% TRSRec 被设计成一个即插即用的模块，可以无缝集成到各种 MSR 骨干网络中。

% 为了确保最大程度的兼容性，TRSRec 直接继承了骨干模型的原始推荐目标 $\mathcal{L}_{\text{rec}}$，无需引入任何额外的训练目标。

% 因此，TRSRec 自然地支持现有 MSR 模型中常用的各种损失函数。例如，$\mathcal{L}_{\text{rec}}$ 可以实例化为二进制交叉熵 (BCE)~\cite{BCE}（SASRec~\cite{SASRec} 中广泛采用的损失函数），或者贝叶斯个性化排序 (BPR)~\cite{BPR}（STOSA~\cite{STOSA} 等模型的核心优化目标）。通过直接复用 MSR 骨干网络的原生训练目标，TRSRec 可以应用于现有和未来的架构，而无需对其原始优化设计进行任何修改。

% 值得注意的是，所提出的TRSRec仅在训练阶段应用。

% 在推理阶段，模型遵循标准的MSR预测流程，不会产生额外的计算或延迟开销。
\subsection{Training and Inference}

REVEAL is designed as a plug-and-play module that can be seamlessly integrated into a wide range of MSR backbones. To ensure maximal compatibility, REVEAL directly inherits the original recommendation objective $\mathcal{L}_{\text{rec}}$ of the backbone model, without introducing any additional training targets. As a result, REVEAL naturally supports various loss functions commonly used in existing MSR models. For example, $\mathcal{L}_{\text{rec}}$ can be instantiated as Binary Cross Entropy (BCE)~\cite{BCE}, which is widely adopted in SASRec~\cite{SASRec}, or Bayesian Personalized Ranking (BPR)~\cite{BPR}, the core optimization objective of models such as STOSA~\cite{STOSA}. By directly reusing the native training objective of the MSR backbone, REVEAL can be applied to both existing and future architectures without any modification to their original optimization design. NNotably, the proposed REVEAL is applied only during training, and its overall training procedure is summarized in Algorithm~\ref{alg:REVEAL}. During inference, the model follows standard MSR pipeline for prediction, incurring no additional computational or latency overhead.

\begin{algorithm}[t]
\caption{Overall Training Procedure of REVEAL}
\label{alg:REVEAL}
\begin{algorithmic}[1]
\Require Item visual content $p^{v}$, ID embedding $e^{id}$, textual embedding $e^{t}$, ground truth $G$, frozen ${\text{VLM}}$, MSR backbone $f_{\text{MSR}}$, evaluation instruction $\text{EvalInstr}$, maximum refinement cycle number $X_{\max}$
\Ensure Trained MSR backbone $f_{\text{MSR}}$ and optimized prompt $\text{Prompt\_X}$

\State Initialize $\text{Prompt\_X}$ \Comment{Extracted from item metadata and user reviews}
\State Set $X \gets 0$ \Comment{Current refinement cycle counter}

\While{$X < X_{\max}$}
    \State \textbf{Step 1: Encoding Visual Content}
    \State $e^{v\_X} \gets {\text{VLM}}(p^{v}, \text{Prompt\_X})$
    
    \State \textbf{Step 2: Training MSR Backbone}
    \State Train $f_{\text{MSR}}$ using $(e^{v\_X}, e^{id}, e^{t})$ under the current prompt
    \State Apply AVL during training to calibrate gradients of vision-related parameters
    
    \State \textbf{Step 3: Recommendation}
    \State $\text{Result\_X} \gets f_{\text{MSR}}(e^{v\_X}, e^{id}, e^{t})$

    \If{$X < X_{\max} - 1$}
        \State \textbf{Step 4: Evaluation and Analysis}
        \State $\text{Critic\_X} \gets {\text{VLM}}\big(\text{Prompt\_X}, \text{EvalInstr}, G, \text{Result\_X}\big)$
        
        \State \textbf{Step 5: Prompt Update}
        \State $\text{Prompt\_(X+1)} \gets {\text{VLM}}(\text{Critic\_X}, \text{Prompt\_X}, p^{v})$
        \State $\text{Prompt\_X} \gets \text{Prompt\_(X+1)}$
    \EndIf
    
    \State $X \gets X + 1$ \Comment{Advance to next refinement cycle}
\EndWhile

\State \Return $f_{\text{MSR}}, \text{Prompt\_X}$
\end{algorithmic}
\end{algorithm}
\section{Experimental Setup}

\subsection{Datasets}
\subsubsection{Datasets Description}
We evaluate REVEAL on four real-world datasets: Amazon Home Beauty, and Sports\footnote{\url{https://snap.stanford.edu/data/amazon/productGraph}} and Yelp\footnote{\url{https://business.yelp.com/data/resources/open-dataset}}. 
The Amazon datasets capture users’ purchasing behaviors and product reviews across multiple domains~\cite{SASRec, STOSA, HM4SR, M3SRec}. 
The Yelp dataset contains user ratings and reviews of local businesses~\cite{Yelp1, Yelp2}. 
All datasets are widely used for sequential recommendation and are preprocessed using the standard 5-core setting~\cite{SASRec, STOSA, Yelp1, Yelp2}, ensuring that each user and item has at least five interactions. 
Detailed statistics are presented in Table~\ref{tab:Statistics}.

\subsubsection{Data Processing}
For all datasets, we first sort user interactions in chronological order to construct behavior sequences. Following the standard 5-core setting, we remove users and items with fewer than five interactions. For each item, we collect its associated multimodal information, including product images and textual descriptions for Amazon, and business photos and textual metadata for Yelp. Items with missing visual or textual information are excluded to ensure consistent multimodal input.

To support sequential recommendation, each user’s historical interactions are organized into a time-ordered sequence, and overly long sequences are truncated to the most recent $L$ interactions, where $L$ is the maximum sequence length used by the backbone model. For multimodal baselines, we keep the textual input construction consistent with their original settings, while visual inputs are further processed in the unified feature extraction pipeline described in Section~\ref{sec:FVE}.

After preprocessing, we adopt the leave-one-out protocol to split each user sequence into training, validation, and test parts, where the last interaction is used for testing, the second last for validation, and the remaining interactions for training.

\begin{table}[!ht]
\centering
\caption{Datasets Statistics}
\begin{tabular}{@{}cccccc@{}}
\toprule
Dataset & \multicolumn{1}{c}{\#users} & \multicolumn{1}{c}{\#items} & \multicolumn{1}{c}{\#interactions} & \multicolumn{1}{c}{density} & \multicolumn{1}{c} {avg.length} \\ 
\midrule
Home & 66,519 & 28,237 & 551,682 & 0.03\% & 8.3 \\
Beauty & 22,363 & 12,101 & 198,502 & 0.06\% & 8.4 \\
Sports& 35,598& 18,358& 296,337& 0.04\%& 8.3\\
Yelp & 287,116 & 148,523 & 4,392,169 & 0.01\% & 15.1\\
\bottomrule
\end{tabular}
\label{tab:Statistics}
\end{table}

\subsection{Baseline Methods}
To evaluate the effectiveness and generality of REVEAL, we compare it with representative baselines from two groups: (1) \textbf{Traditional Recommendation:} Methods based on item co-occurrence modeling, including SASRec~\cite{SASRec}, BERT4Rec~\cite{Bert4rec}, LightGCN~\cite{LightGCN}, and STOSA~\cite{STOSA}; (2) \textbf{Multimodal Recommendation:} Methods incorporating additional modalities (e.g., vision and text), including VBPR~\cite{VBPR}, UniSRec~\cite{UniSRec}, MMSBR~\cite{MMSBR}, MISSRec~\cite{MISSRec}, M3SRec~\cite{M3SRec}, and HM4SR~\cite{HM4SR}.

\textbf{(1) Traditional Recommendation:}  

\begin{itemize}
\item SASRec (ICDM 2018)\footnote{\url{https://github.com/kang205/SASRec}}~\cite{SASRec}: A self-attention based sequential recommendation model that predicts the next item by selecting the most relevant interactions from a user’s history, combining the strengths of Markov Chains and RNNs.
\item BERT4Rec (CIKM 2019)\footnote{\url{https://github.com/FeiSun/BERT4Rec}}~\cite{Bert4rec}: A bidirectional Transformer-based sequential recommendation model that learns user representations by predicting masked items from both left and right context, enabling more expressive sequence modeling than unidirectional methods.
\item LightGCN (SIGIR 2020)\footnote{\url{https://github.com/enoche/MMRec}}~\cite{LightGCN}: A simplified graph-based collaborative filtering model that learns user and item embeddings by linearly propagating them over the user-item interaction graph, retaining only neighborhood aggregation and removing feature transformation and nonlinear activation.
\item STOSA (WWW 2022)\footnote{\url{https://github.com/zfan20/STOSA}}~\cite{STOSA}: A stochastic self-attention based sequential recommendation model that represents items as Gaussian distributions and uses Wasserstein self-attention to capture uncertainty and item transitions in user behavior sequences.
\end{itemize}

\textbf{(2) Multimodal Recommendation:}  

\begin{itemize}
\item VBPR (AAAI 2016)\footnote{\url{https://github.com/enoche/MMRec}}~\cite{VBPR}: A visually-aware personalized ranking model that incorporates image features extracted by pre-trained deep networks into matrix factorization, improving recommendation accuracy and alleviating cold-start issues.
\item UniSRec (KDD 2022)\footnote{\url{https://github.com/RUCAIBox/UniSRec}}~\cite{UniSRec}: A universal sequential recommendation model that learns transferable item and sequence representations from item textual descriptions via multi-domain contrastive pre-training, enabling efficient adaptation to new domains without relying on item ID overlap.
\item MMSR (CIKM 2023)\footnote{\url{https://github.com/HoldenHu/MMSR}}~\cite{MMSR}: A graph-based multimodal sequential recommendation model that represents each user’s history as a cross-modal interaction graph and adaptively determines the fusion order of modality features through gated graph propagation.
\item MMSBR (TKDE 2023)\footnote{\url{https://github.com/Zhang-xiaokun/MMSBR}}~\cite{MMSBR}: A multimodal session-based recommendation model that unifies descriptive and numerical item information, using contrastive learning, hierarchical fusion, and Wasserstein self-attention to better infer user intent from short sessions.
\item MISSRec (MM 2023)\footnote{\url{https://github.com/gimpong/MM23-MISSRec}}~\cite{MISSRec}: A multimodal sequential recommendation framework that leverages multimodal pre-training and user-adaptive item fusion to learn robust and transferable sequence representations beyond sparse ID features.
\item M3SRec (CIKM 2023)~\cite{M3SRec}: A multimodal sequential recommendation model that uses a mixture-of-experts fusion network and task-aligned pre-training to jointly reduce the semantic gap across modalities and improve user preference modeling under sparse data.
\item HM4SR (WWW 2025)\footnote{\url{https://github.com/SStarCCat/HM4SR}}~\cite{HM4SR}: A multimodal sequential recommendation model that uses a hierarchical mixture-of-experts framework to filter interest-irrelevant multimodal noise and capture dynamic user preferences with explicit temporal information.
\end{itemize}

% \subsection{Evaluation Settings}
% We adopt the standard leave-one-out protocol for dataset splitting~\cite{SASRec, STOSA}. Performance is evaluated using Recall@K (R@K) and NDCG@K (N@K) with $K \in \{10, 20\}$~\cite{SASRec, Bert4rec, LightGCN, STOSA}, where higher values indicate better performance.
\subsection{Evaluation Settings}
We adopt the standard leave-one-out protocol for dataset splitting~\cite{SASRec, STOSA}. For each user sequence, the last interaction is used for testing, the second last interaction for validation, and the remaining interactions for training. Performance is evaluated using Recall@K (R@K) and NDCG@K (N@K) with $K \in \{10, 20\}$~\cite{SASRec, Bert4rec, LightGCN, STOSA}, where higher values indicate better performance. Following the standard full-ranking protocol, each ground-truth test item is ranked against all candidate items that do not appear in the user’s training sequence. The same evaluation setting is used for all compared methods. All results are averaged over five runs with different random seeds. Statistical significance is assessed using a paired two-sided t-test against the strongest baseline, and improvements are considered significant when $p < 0.05$. Detailed RecBole evaluation configurations are provided in the supplementary material.

\subsection{Implementation Details}
All models are implemented in PyTorch based on the open-source RecBole framework~\cite{RecBole}. For fair comparison, baseline models are configured using the best hyperparameters reported in their original papers or tuned via grid search when such hyperparameters are not available.

Unless otherwise specified, REVEAL uses Qwen2.5-VL-7B as the default frozen VLM backbone in all stages of PVE, including vision encoding, prompt analysis, and prompt refinement. We choose Qwen2.5-VL-7B because it provides a practical balance between multimodal understanding capability and computational efficiency, while naturally supporting both visual feature extraction and language-based prompt optimization within a unified framework.

For visual representation extraction, each item image is fed into the frozen Qwen2.5-VL-7B together with the current prompt, and the resulting image-aware hidden representation is used as the visual embedding. The extracted feature is then projected into the input dimensionality required by each MSR backbone through a trainable linear layer, enabling PVE to be integrated into different recommendation backbones without changing their original architectures.

We do not adopt CLIP as the visual encoder because CLIP mainly provides static image-text alignment features and cannot support the iterative prompt analysis and refinement process required by PVE. We also do not use embedding-oriented VLMs such as Qwen3-VL-Embedding, because PVE requires not only visual embedding extraction but also language generation for critic-based prompt updating. In contrast, a generative VLM can jointly support vision encoding, textual evaluation, and prompt rewriting within a unified framework.

During training, the VLM parameters remain frozen, while the prompt is iteratively updated through the three-stage PVE pipeline, including Vision Encoding, Recommendation, and Prompt Analysis. In our implementation, prompt optimization is performed at a coarser granularity than backbone parameter updates. Specifically, under a fixed prompt, the MSR backbone is first trained to convergence using visual embeddings extracted in the Vision Encoding stage. The resulting recommendation outputs are then analyzed in the Prompt Analysis stage to assess prompt quality and refine the prompt for the next cycle. Therefore, prompt refinement is conducted only after each complete backbone training stage, rather than at every batch or epoch. This alternating training-and-refinement process is repeated for a fixed number of refinement cycles $X_{\max}$, unless otherwise specified.

For prompt refinement, we use deterministic decoding with a fixed random seed to improve reproducibility. Unless otherwise stated, greedy decoding is adopted for textual critic generation and prompt updating. Since REVEAL does not maintain separate prompts for individual users, the additional storage overhead of PVE is negligible.

For textual representations, we follow the original input setting of each baseline whenever textual features are required. REVEAL only modifies the visual representation extraction stage through PVE and the optimization stage through AVL, while keeping the sequence modeling backbone unchanged. In particular, for multimodal baselines such as MISSRec, M3SRec, and HM4SR, we replace their original visual encoders with Qwen2.5-VL-7B to ensure compatibility with prompt-guided preference-aware visual extraction.

For REVEAL, the For REVEAL, the hyperparameters mainly include the calibration strength parameter $\alpha$ in AVL and the maximum prompt refinement cycle number $X_{\max}$ in PVE. We select $\alpha$ on the validation set through grid search over $\{0.01, 0.05, 0.1, 0.2, 0.5, 1.0\}$, and tune $X_{\max}$ over $\{1,2,3,4,5\}$. All results are averaged over five runs with different random seeds. Statistical significance is assessed using a paired two-sided t-test, and improvements are considered significant when $p < 0.05$. All experiments are conducted on 8 $\times$ 4090 GPU.
\section{Experiment}
To provide a comprehensive evaluation of REVEAL, we conduct experiments on real-world public datasets from multiple perspectives, including its overall recommendation effectiveness, the contribution of visual features, the effectiveness of feedback-guided visual extraction, the effects of VLM backbone scale and selection, the sensitivity of adaptive visual learning to hyperparameter settings, and the computational complexity of the framework. Specifically, we aim to answer the following research questions:
\begin{itemize}
    \item \textbf{RQ1:} What is the overall effectiveness of REVEAL?
    \item \textbf{RQ2:} How do visual features contribute to REVEAL?
    \item \textbf{RQ3:} How does FVE improve the extraction of visual information?
    \item \textbf{RQ4:} What are the effects of VLM backbone scale and selection on FVE?
    \item \textbf{RQ5:} How sensitive is AVL to different hyperparameter settings?
    \item \textbf{RQ6:} What additional computational cost does REVEAL introduce?
\end{itemize}

%Baseline Datasets implentment details
%放表 分析

\begin{table*}[!ht]
    \small
    \setlength\tabcolsep{2pt}
    \caption{Performance comparison against all baselines. 
    \textit{w/} REVEAL denotes the incorporation of our framework into the corresponding backbone. Improve. reports the relative gain over the vanilla model. $\dagger$ indicates results reproduced by ourselves. All results are averaged over five runs and are statistically significant ($p < 0.05$).}
    \label{tab:compare}
    \resizebox{0.96\linewidth}{!}{
        \begin{tabular}{c|cccc|cccc|cccc|cccc}
        \hline
        \makebox[0.15\linewidth]{Datasets}        & \multicolumn{4}{c|}{Home} & \multicolumn{4}{c|}{Beauty}  & \multicolumn{4}{c|}{Sports}    & \multicolumn{4}{c}{Yelp}                                                                             \\ \hline
        Metric                                    & {R@10}& {R@20}& {N@10}&{N@20}                      & {R@10}& {R@20}& {N@10}& {N@20}                      
        &{R@10}& {R@20}& {N@10}& {N@20}                       & {R@10}& {R@20}& {N@10}& {N@20}  \\ \hline
        
        \rowcolor{gray!10}\multicolumn{17}{c}{\footnotesize Traditional Recommendation} \\ \hline
        SASRec                          & 0.0168& 0.0249& 0.0081& 0.0099                     & 0.0534& 0.0839& 0.0247& 0.0332                      
        &0.0336& 0.0505& 0.0175& 0.0218           & 0.0233& 0.0391& 0.0123& 0.0152  \\
        BERT4Rec                         & 0.0156& 0.0238& 0.0073& 0.0101                     & 0.0545& 0.0852& 0.0254& 0.0351                      
        &0.0342& 0.0512& 0.0172& 0.0216           & 0.0245& 0.0411& 0.0119& 0.0162  \\ 
        LightGCN                        & 0.0161& 0.0243& 0.0077& 0.0103                     & 0.0549& 0.0861& 0.0258& 0.0355                      
        &0.0366& 0.0535& 0.0203& 0.0242           & 0.0236& 0.0414& 0.0123& 0.0166  \\
        STOSA                             & 0.0169& 0.0264& 0.0098& 0.0113                     & 0.0648& 0.0941& 0.0339& 0.0385                      
        &0.0389& 0.0560& 0.0220& 0.0264           & 0.0238& 0.0424& 0.0128& 0.0161  \\\hline
        
        \rowcolor{gray!10}\multicolumn{17}{c}{\footnotesize Multimodal Recommendation} \\ \hline
        VBPR                             & 0.0159& 0.0256& 0.0074& 0.0095                     & 0.0551& 0.0842& 0.0252& 0.0337                      
        &0.0321& 0.0492& 0.0159& 0.0195           & 0.0237& 0.0434& 0.0123& 0.0163  \\
        UniSRec                           & 0.0198& 0.0286& 0.0121& 0.0129                     & 0.0661& 0.0988& 0.0386& 0.0439                      
        &0.0361& 0.0544&0.0199&0.0223             & 0.0246& 0.0442& 0.0137& 0.0167  \\
        MMSR                             & 0.0218 & 0.0322& 0.0135& 0.0142                     
        & 0.0715& 0.1058& 0.0401& 0.0446                      
        & 0.0378& 0.0567& 0.0223& 0.0237          & 0.0245 & 0.0456 & 0.0129& 0.0156  \\
        MMSBR                            & 0.0238& 0.0331& 0.0142& 0.0152                    & 0.0719& 0.1062&0.0406&0.0474 
        & 0.0393& 0.0579& 0.0235& 0.0245          &0.0255& 0.0464&0.0139& 0.0166  \\\hline
        
        \rowcolor{gray!10}\multicolumn{17}{c}{\footnotesize Multimodal Sequential Recommendation} \\ \hline
        MISSRec                             & 0.0227& 0.0310& 0.0148& 0.0169         
            &0.0689& 0.0988&0.0407&0.0473         
        & 0.0378& 0.0529& 0.0233& 0.0271          &0.0259& 0.0473&0.0141&0.0161   \\
        \rowcolor{gray!30}\  \textit{w/} REVEAL              &0.0239& 0.0328&0.0167&0.0187
        &0.0730& 0.1037& 0.0435&0.0504 
        &0.0397& 0.0550& 0.0252& 0.0285                     &0.0268& 0.0486& 0.0152&0.0170 \\
        Improve.                                  &5.29\%& 5.81\%&12.84\%&10.65\% 
        &5.95\%&4.96\%&6.88\%&6.55\%
        &5.03\%&3.97\%&8.15\%&5.17\%                        &3.47\%& 2.75\%&7.80\%&5.59\% \\\hline
        
        M3SRec $\dagger$                            & 0.0237& 0.0338& 0.0148& 0.0174                    
        &0.0712& 0.1088& 0.0439& 0.0501        
        &0.0410& 0.0568& 0.0242& 0.0286            &0.0251& 0.0494& 0.0155& 0.0171  \\   
        \rowcolor{gray!30}\  \textit{w/} REVEAL      &0.0251& 0.0355&0.0160&0.0188 
        &0.0739& 0.1137&0.0457&0.0522 
        &0.0434& 0.0599&0.0258&0.0303              &0.0262& 0.0514&0.0168&0.0184 \\
        Improve.                                    &5.91\%&5.03\%&8.11\%&8.05\%  
        &3.79\%&4.50\%&4.10\%&4.19\% 
        &5.85\%&5.46\%&6.61\%&5.94\%                        &4.38\%&4.05\%&8.39\%&7.60\%\\\hline   
        
        HM4SR                              & 0.0323& 0.0449& 0.0188& 0.0217                    
        & 0.0921& 0.1271& 0.0555& 0.0640 
        & 0.0515& 0.0744& 0.0289& 0.0348           & 0.0374& 0.0563& 0.0249& 0.0286\\ 
        \rowcolor{gray!30}\ \textit{w/} REVEAL       &0.0341& 0.0478&0.0205&0.0236
        &0.0962& 0.1342&0.0587&0.0675                         
        &0.0548& 0.0786&0.0313&0.0371              &0.0393& 0.0591&0.0263&0.0301 \\
        Improve.                                   &5.57\%&6.46\%&9.04\%&8.76\%
        &4.45\%&5.59\%&5.77\%&5.47\%
        &6.41\%&5.65\%&8.30\%&6.61\%                        &5.08\%&4.97\%&5.62\%&5.24\% \\\hline
        \end{tabular}
        }
\end{table*}

\subsection{Overall Performance Comparison (RQ1)}
To comprehensively evaluate the effectiveness and generalizability of REVEAL, we compare its enhanced variants with a broad range of baselines across four datasets. The results in Table~\ref{tab:compare} lead to the following observations.

(1) REVEAL consistently improves all three MSR backbones across all datasets and metrics, which suggests that its effectiveness is not tied to a specific architecture. This result is important because it indicates that the gains mainly come from the proposed visual enhancement framework rather than from backbone-specific design. For example, when applied to MISSRec on Sports, REVEAL improves NDCG@10 by 8.15\% and NDCG@20 by 5.17\%. Similar improvements are also observed on M3SRec and HM4SR, demonstrating that REVEAL can serve as a robust plug-and-play framework for different MSR models.

(2) The performance gains are generally more notable on ranking-based metrics such as NDCG, which indicates that REVEAL not only helps retrieve relevant items but also improves their positions in the ranked list. This trend suggests that the proposed framework enhances the quality of visual representations in a way that is more aligned with user preference modeling. For instance, on Home, REVEAL improves MISSRec by 5.29\% in Recall@10 but by 12.84\% in NDCG@10. Such results imply that preference-relevant visual cues are better captured and utilized, leading to more accurate top-ranked recommendations.

(3) The improvement margins vary across datasets, which further suggests that the contribution of visual signals is dataset-dependent. On datasets such as Home and Sports, the gains are relatively larger, indicating that these datasets likely contain more useful but underexploited visual information. In contrast, the improvements on Yelp are relatively smaller in some settings, implying that visual cues may play a less dominant role in this domain. Nevertheless, REVEAL still produces consistent gains on Yelp, which shows that enhancing visual extraction and visual learning remains beneficial even when the visual modality is not the primary signal source.

\subsection{Ablation Study (RQ2)}
We investigate the effectiveness of the proposed FVE and AVL modules with an ablation study. The study was conducted under two settings: a multimodal setting utilizing both textual and visual features, and an vision-only setting.
The study was conducted on two MSR backbones: MISSRec, M3SRec and HM4SR.

% 还差一个消融模态的部分
\subsubsection{The Multimodal Setting}

As shown in Table~\ref{table:ablation}, the ablation results in the multimodal setting lead to the following observations:

(1) Both FVE and AVL consistently improve the base models across all three MSR backbones, indicating that each module contributes positively to multimodal recommendation. This consistency is important because it suggests that the gains are not caused by a specific backbone, but by the improved use of visual information. In particular, \textit{w/} FVE achieves stable improvements on all models, which indicates that preference-aware visual extraction helps derive more recommendation-relevant visual cues from item images. Similarly, \textit{w/} AVL also improves all backbones, suggesting that adaptively strengthening visual learning during optimization can enhance the contribution of visual signals.

(2) Compared with AVL, FVE generally brings slightly larger gains on most metrics, which suggests that improving visual feature quality at the extraction stage may have a stronger effect than optimization alone. For example, on HM4SR, \textit{w/} FVE improves Recall@20 from 0.0449 to 0.0467, while \textit{w/} AVL improves it to 0.0461. A similar trend can also be observed on MISSRec and M3SRec. This result indicates that if the extracted visual representations are more aligned with user preferences at the beginning, the downstream recommendation model can benefit more directly.

(3) The full REVEAL framework consistently achieves the best performance across all metrics and backbones, showing that FVE and AVL are complementary rather than redundant. When the two modules are combined, the improvements become more pronounced than using either module alone. For instance, on HM4SR, REVEAL improves Recall@20 from 0.0449 to 0.0478 and NDCG@20 from 0.0217 to 0.0236, outperforming both single-module variants. This suggests that better visual extraction and better visual learning address different parts of the visual utilization problem, and their combination enables more effective multimodal recommendation.

\begin{table}[t]
    \caption{Ablation results of different variants on Home dataset. ``\textit{w/}'' denotes ``with.'' The Best results are in \textbf{bold}. All results are averaged over five runs with $p < 0.05$.}
    \label{table:ablation}
    \centering
    \begin{tabular}{llcccc}
        \toprule
        &\textbf{Variant}&  \textbf{Recall@10} & \textbf{Recall@20} & \textbf{NDCG@10}& \textbf{NDCG@20} \\
        \midrule
        \multirow{4}{*}{MISSRec}
        &Base Model          & 0.0227 & 0.0310 & 0.0148 & 0.0169\\ 
        &\textit{w/} FVE & 0.0234 & 0.0325 & 0.0153 & 0.0177 \\
        &\textit{w/} AVL & 0.0231 & 0.0321 & 0.0150 & 0.0173 \\
        &\textit{w/} REVEAL & \textbf{0.0239} & \textbf{0.0328} & \textbf{0.0167} & \textbf{0.0187}\\\hline
        \multirow{4}{*}{M3SRec}
        &Base Model          & 0.0237 & 0.0338 & 0.0148 & 0.0174\\ 
        &\textit{w/} FVE & 0.0245 & 0.0348 & 0.0156 & 0.0184 \\
        &\textit{w/} AVL & 0.0241 & 0.0345 & 0.0152 & 0.0180 \\
        &\textit{w/} REVEAL & \textbf{0.0251} & \textbf{0.0355} & \textbf{0.0160} & \textbf{0.0188}\\\hline
        \multirow{4}{*}{HM4SR} 
        &Base Model           & 0.0323 & 0.0449 & 0.0188 & 0.0217\\ 
        &\textit{w/} FVE & 0.0336 & 0.0467 & 0.0200 & 0.0230 \\
        &\textit{w/} AVL & 0.0332 & 0.0461 & 0.0197 & 0.0228 \\
        &\textit{w/} REVEAL  & \textbf{0.0341} & \textbf{0.0478} & \textbf{0.0205} & \textbf{0.0236}\\ 
        \bottomrule
    \end{tabular}
\end{table}

\begin{table}[htbp]
\caption{Performance comparison under the Vision-Only (V-Only) setting across datasets and variants, where text embeddings are replaced with zero vectors. The Best results are in \textbf{bold}. All Results are averaged over five runs with $p < 0.05$.}
\label{tab:Modality_ablation}
\resizebox{0.8\linewidth}{!}{
\centering
\begin{tabular}{cccccccc}
\toprule
\textbf{Dataset} & \textbf{Variant} & \textbf{Recall@10} & \textbf{Recall@20} & \textbf{NDCG@10} & \textbf{NDCG@20} \\
\midrule
\multirow{4}{*}{Home}
& V-Only & 0.0202 & 0.0287 & 0.0126 & 0.0148 \\
& V-Only + FVE & 0.0208 & 0.0296 & 0.0130 & 0.0152 \\
& V-Only + AVL & 0.0210 & 0.0298 & 0.0131 & 0.0154 \\
& V-Only + REVEAL & \textbf{0.0218} & \textbf{0.0310} & \textbf{0.0136} & \textbf{0.0160} \\
\midrule
\multirow{4}{*}{Beauty}
& V-Only & 0.0605 & 0.0925 & 0.0373 & 0.0426 \\
& V-Only + FVE & 0.0623 & 0.0953 & 0.0384 & 0.0439 \\
& V-Only + AVL & 0.0629 & 0.0962 & 0.0388 & 0.0443 \\
& V-Only + REVEAL & \textbf{0.0643} & \textbf{0.0989} & \textbf{0.0403} & \textbf{0.0460} \\
\midrule
\multirow{4}{*}{Sports}
& V-Only & 0.0349 & 0.0483 & 0.0206 & 0.0243 \\
& V-Only + FVE & 0.0359 & 0.0497 & 0.0212 & 0.0250 \\
& V-Only + AVL & 0.0363 & 0.0502 & 0.0214 & 0.0253 \\
& V-Only + REVEAL & \textbf{0.0377} & \textbf{0.0522} & \textbf{0.0222} & \textbf{0.0262} \\
\midrule
\multirow{4}{*}{Yelp}
& V-Only & 0.0213 & 0.0420 & 0.0132 & 0.0145 \\
& V-Only + FVE & 0.0219 & 0.0433 & 0.0136 & 0.0149 \\
& V-Only + AVL & 0.0222 & 0.0437 & 0.0137 & 0.0151 \\
& V-Only + REVEAL & \textbf{0.0230} & \textbf{0.0454} & \textbf{0.0143} & \textbf{0.0157} \\
\bottomrule
\end{tabular}
}
\end{table}

\subsubsection{The Vision-Only Setting}

To further examine the effectiveness of FVE and AVL when textual information is unavailable, we evaluate M3SRec+REVEAL under the Vision-Only setting across all datasets by replacing text embeddings with zero vectors. The results in Table~\ref{tab:Modality_ablation} lead to the following observations.

(1) Removing the text modality consistently degrades performance across all datasets compared with the full multimodal setting, indicating that visual information alone is not sufficient to fully support accurate recommendation. This result suggests that text embeddings still provide important semantic cues that cannot be completely replaced by visual signals. The degradation is particularly noticeable on datasets such as Beauty and Sports, which implies that multimodal recommendation benefits from the complementary interaction between text and vision rather than relying on a single modality alone.

(2) Under the Vision-Only setting, both FVE and AVL still bring consistent improvements over the V-Only baseline on all datasets, showing that the proposed modules remain effective even when visual information becomes the primary content signal. This is important because it suggests that the gains of REVEAL do not depend on the presence of text, but are directly related to better visual modeling. For example, on the Beauty dataset, Recall@20 increases from 0.0925 to 0.0953 with FVE and to 0.0962 with AVL. Similar trends can also be observed on Home, Sports, and Yelp. These results indicate that FVE helps derive more preference-relevant visual features, while AVL improves how visual information is optimized and utilized during training.

(3) The full REVEAL framework consistently achieves the best performance across all datasets and metrics, which suggests that FVE and AVL remain complementary in the absence of textual input. Compared with using either module alone, their combination yields larger and more stable gains. For instance, on Sports, Recall@20 improves from 0.0483 in the V-Only baseline to 0.0497 with FVE and 0.0502 with AVL, while the full REVEAL further raises it to 0.0522. This result indicates that improving visual feature extraction and strengthening visual learning address different aspects of the visual underutilization problem, and their joint use is more effective under vision-dominant conditions.

\subsection{Visual Analysis of Case Study (RQ3)}
%放一下注意力可视化图片然后进行分析
\begin{figure*}
    \centering
    \includegraphics[width=0.96\linewidth]{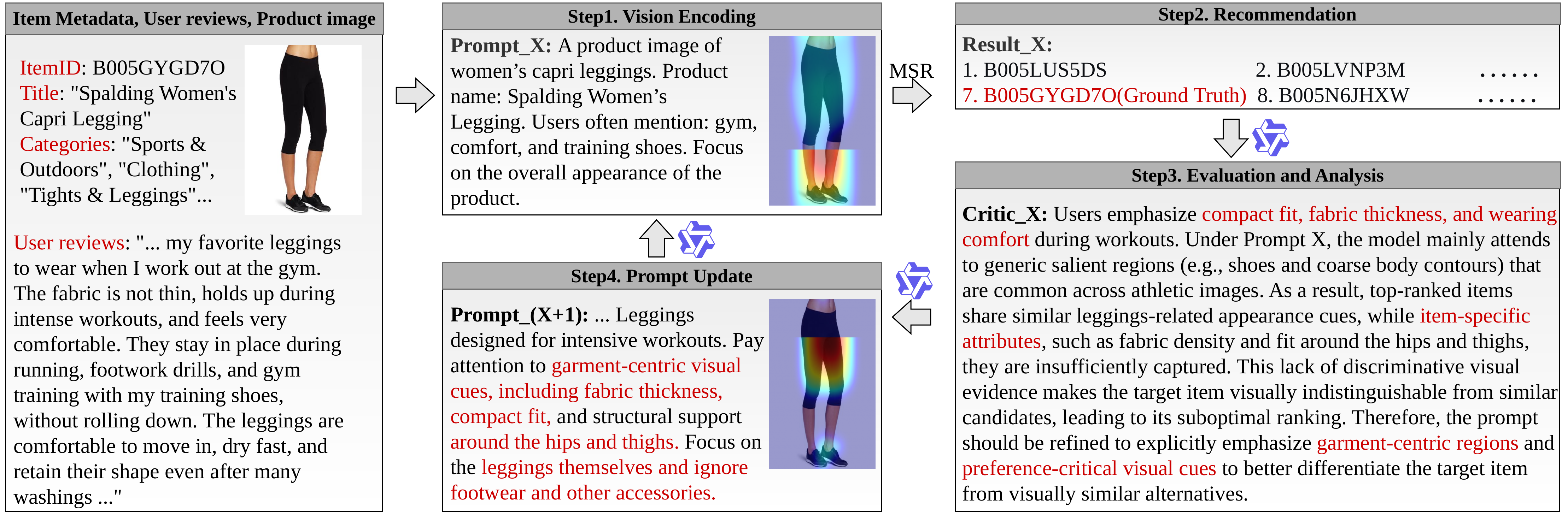}
    \caption{Case study illustrating how FVE refines visual features extraction through prompt optimization on the Sports dataset.}
    \Description{The Case Study Figure.}
    \label{fig:Case Study}
\end{figure*}

To further examine how FVE improves recommendation through feedback-guided prompt refinement, we conduct a case study on a representative item from the Sports dataset. As shown in Figure~\ref{fig:Case Study}, the initial prompt is constructed from item metadata and review keywords, and is then used to guide Qwen2.5-VL-7B for visual feature extraction. We visualize cross-modal interactions using attention heatmaps derived from the final-layer attention of the last query token over visual patch tokens, averaged across heads and normalized to highlight the image regions emphasized by the textual context. The analysis leads to the following observations.

(1) Under the initial prompt $\text{Prompt\_X}$, the visual attention is biased toward generic salient regions rather than recommendation-critical item attributes. In this case, the prompt emphasizes the overall appearance of the product and includes review terms such as ``gym,'' ``comfort,'' and ``training shoes.'' As a result, the model attends strongly to visually prominent but weakly discriminative regions, especially footwear and coarse body contours, instead of focusing on the leggings themselves. This behavior is important because such regions are easily shared across many sportswear images and therefore contribute limited evidence for distinguishing the target item from visually similar candidates. Consequently, although the extracted representation captures general athletic context, it fails to encode the fine-grained garment cues that are more directly related to user preference. :contentReference[oaicite:1]{index=1}

(2) The ranking error suggests that visually salient regions are not necessarily the most recommendation-relevant regions. As shown in Figure~\ref{fig:Case Study}, the ground-truth item is ranked only seventh under $\text{Prompt\_X}$. The critic analysis indicates that top-ranked items share broadly similar leggings-related appearance, while item-specific attributes, such as fabric density and fit around the hips and thighs, are insufficiently captured. This result is analytically meaningful because it reveals a key limitation of static prompt-guided extraction: if the prompt remains at the level of coarse appearance, the resulting visual representation may favor common patterns over discriminative preference cues. In other words, the model is not simply attending to the ``wrong'' region in a generic sense; rather, it is attending to regions that are visually obvious but insufficiently informative for recommendation ranking. :contentReference[oaicite:2]{index=2}

(3) After feedback-guided refinement, attention shifts toward garment-centric and preference-relevant regions, which suggests that FVE can use recommendation errors to improve visual extraction without updating VLM parameters. The refined prompt $\text{Prompt\_(X+1)}$ explicitly emphasizes fabric thickness, compact fit, and structural support around the hips and thighs, while suppressing irrelevant cues such as footwear and accessories. Correspondingly, the heatmap shows a clearer concentration on leggings-specific regions rather than generic sportswear context. This shift indicates that the prompt update helps align visual attention with the actual source of ranking discrimination. Notably, the refinement does not change the VLM itself; instead, it changes how the frozen VLM interprets the same image under recommendation feedback. Therefore, this case study provides qualitative evidence that FVE improves recommendation not by increasing model capacity, but by making visual extraction more preference-aware and more consistent with the ranking objective. :contentReference[oaicite:3]{index=3}.

\subsection{Effect of VLM Backbone Scale and Selection (RQ4)}

\subsubsection{Effect of VLM Backbone Scale}

\begin{table}[t]
    \caption{Performance comparison of FVE with Qwen2.5-VL(Q-VL) across scaling sizes. The best results are highlighted in \textbf{bold}. All results are averaged over five runs.}
    \label{table:model_scale}
    \centering
    \resizebox{0.8\linewidth}{!}{
    \begin{tabular}{llcccc}
        \toprule
        \textbf{Dataset} & \textbf{Model} & \textbf{Recall@10} & \textbf{Recall@20} & \textbf{NDCG@10} & \textbf{NDCG@20} \\
        \midrule
        \multirow{3}{*}{Home}
        & Qwen2.5-VL-3B  & 0.0248 & 0.0349 & 0.0158 & 0.0184 \\
        & Qwen2.5-VL-7B  & 0.0251 & 0.0355 & 0.0160 & 0.0188 \\
        & Qwen2.5-VL-32B & \textbf{0.0252} & \textbf{0.0359} & \textbf{0.0164} & \textbf{0.0190} \\
        \midrule
        \multirow{3}{*}{Beauty}
        & Qwen2.5-VL-3B  & \textbf{0.0748} & 0.1134 & \textbf{0.0458} & 0.0509 \\
        & Qwen2.5-VL-7B  & 0.0739 & \textbf{0.1137} & 0.0457 & 0.0522 \\
        & Qwen2.5-VL-32B & 0.0744 & 0.1135 & 0.0456 & \textbf{0.0533} \\
        \midrule
        \multirow{3}{*}{Sports}
        & Qwen2.5-VL-3B  & 0.0426 & 0.0593 & 0.0252 & 0.0300 \\
        & Qwen2.5-VL-7B  & \textbf{0.0434} & 0.0599 & 0.0258 & \textbf{0.0303} \\
        & Qwen2.5-VL-32B & 0.0431 & \textbf{0.0602} & \textbf{0.0260} & 0.0301 \\
        \midrule
        \multirow{3}{*}{Yelp}
        & Qwen2.5-VL-3B & 0.0255 & 0.0509 & 0.0166 & 0.0175 \\
        & Qwen2.5-VL-7B & 0.0262 & 0.0514 & 0.0168 & \textbf{0.0184} \\
        & Qwen2.5-VL-32B & \textbf{0.0265} & \textbf{0.0519} & \textbf{0.0170} & 0.0183 \\
        \bottomrule
    \end{tabular}
    }
\end{table}

To investigate how the scale of the VLM backbone affects FVE and the final recommendation performance, we compare three Qwen2.5-VL variants, namely 3B, 7B, and 32B. The results in Table~\ref{table:model_scale} lead to the following observations.

(1) Increasing the VLM scale generally brings performance gains, but the improvements are modest rather than dramatic. This suggests that a stronger VLM can provide better visual understanding for FVE, thereby helping derive more informative visual representations, but the benefit is not unlimited. For example, the 32B model achieves the best results on most metrics in Home and Yelp. This trend indicates that larger VLMs can improve visual feature quality, yet the resulting recommendation gains remain relatively constrained.

(2) The effect of model scale is not strictly monotonic across datasets and metrics, which suggests that larger VLMs do not always translate into better recommendation performance. On Beauty, the 3B model performs best on Recall@10 and NDCG@10, while the 32B model only shows an advantage on NDCG@20. On Sports, the 7B model achieves the best Recall@10 and NDCG@20, whereas the 32B model performs best on Recall@20 and NDCG@10. These results indicate that the relationship between VLM capacity and recommendation quality is influenced by dataset characteristics.

(3) The overall results suggest that selecting VLM scale should balance effectiveness and efficiency, rather than assuming that larger models are always preferable. Since 7B and 32B achieve similar performance in many settings, the additional computational cost of larger backbones may not always be justified. Therefore, in practical applications, an appropriate VLM scale should be chosen according to the dataset and the available resource budget. This also suggests that the effectiveness of FVE depends not only on backbone strength, but also on how well the extracted visual signals match the recommendation task itself.

\subsubsection{Effect of VLM Backbone Selection}

\begin{table*}[t]
    \caption{Performance comparison of FVE across different VLM backbones. The best results are highlighted in \textbf{bold}. $\dagger$ indicates the baseline setting of our method, where Qwen2.5-VL-7B is used as the default VLM backbone. All results are averaged over five runs.}
    \label{table:model_backbone}
    \centering
    \resizebox{0.8\linewidth}{!}{
    \begin{tabular}{llcccc}
        \toprule
        \textbf{Dataset} & \textbf{Model} & \textbf{Recall@10} & \textbf{Recall@20} & \textbf{NDCG@10} & \textbf{NDCG@20} \\
        \midrule
        \multirow{10}{*}{Home}
        & Qwen2.5-VL-7B $\dagger$  & 0.0251 & 0.0355 & 0.0160 & 0.0188 \\\cmidrule(lr){2-6}
        & Gemma-3-4B      & 0.0240 & 0.0340 & 0.0153 & 0.0180 \\
        & Gemma-3-12B     & 0.0247 & 0.0349 & 0.0158 & 0.0185 \\
        & Gemma-3-27B     & 0.0254 & 0.0361 & 0.0163 & 0.0191 \\\cmidrule(lr){2-6}
        & InternVL3-8B    & 0.0248 & 0.0351 & 0.0159 & 0.0186 \\
        & InternVL3-14B   & 0.0256 & 0.0362 & 0.0164 & 0.0192 \\
        & InternVL3-38B   & 0.0262 & 0.0370 & 0.0168 & 0.0197 \\\cmidrule(lr){2-6}
        & Qwen3-VL-4B     & 0.0245 & 0.0346 & 0.0156 & 0.0183 \\
        & Qwen3-VL-8B     & 0.0258 & 0.0364 & 0.0165 & 0.0193 \\
        & Qwen3-VL-32B    & \textbf{0.0265} & \textbf{0.0374} & \textbf{0.0170} & \textbf{0.0199} \\
        \midrule
    \end{tabular}}
\end{table*}

Table~\ref{table:model_backbone} reports the performance of FVE with different VLM backbones on the Home dataset. Overall, the results show that the choice of VLM backbone has a noticeable impact on recommendation performance, indicating that the visual understanding capability of the backbone directly affects the effectiveness of feedback-guided visual extraction.

(1) A clear performance improvement can be observed as model size increases within each VLM family. For Gemma-3, Recall@20 improves from 0.0340 to 0.0361 and NDCG@20 from 0.0180 to 0.0191 when scaling from 4B to 27B. A similar trend is observed in the InternVL3 series, where larger backbones consistently yield better performance, and InternVL3-38B reaches 0.0370 in Recall@20 and 0.0197 in NDCG@20. The Qwen3-VL family also shows consistent gains with increasing model size, and Qwen3-VL-32B achieves the best overall performance. These results suggest that larger VLMs provide stronger multimodal representations, which are more beneficial for extracting user-relevant visual signals.

(2) Most larger backbones outperform the default backbone, Qwen2.5-VL-7B. For example, Gemma-3-27B, InternVL3-14B, InternVL3-38B, Qwen3-VL-8B, and Qwen3-VL-32B all surpass the baseline on all four metrics. Among them, Qwen3-VL-32B achieves the best results, improving Recall@20 from 0.0355 to 0.0374 and NDCG@20 from 0.0188 to 0.0199. In contrast, smaller models such as Gemma-3-4B and Qwen3-VL-4B underperform the baseline, suggesting that limited visual modeling capacity constrains the quality of extracted visual features.

(3) These results verify that FVE is compatible with different VLM backbones and can consistently benefit from stronger visual encoders. More importantly, the superior performance of larger backbones indicates that enhancing the underlying visual-language modeling capability is a promising direction for further improving feedback-guided visual extraction.

\subsection{Hyperparameter Analysis (RQ5)}\label{sec:hyperparam}

To further examine the robustness of REVEAL, we conduct a sensitivity analysis of M3SRec+REVEAL with respect to key hyperparameters across all datasets.

\subsubsection{Effect of the Refinement Cycle Number $X_{\max}$}

\begin{figure*}[t]
    \centering
    \begin{subfigure}[t]{0.24\textwidth}
        \centering
        \includegraphics[width=\linewidth]{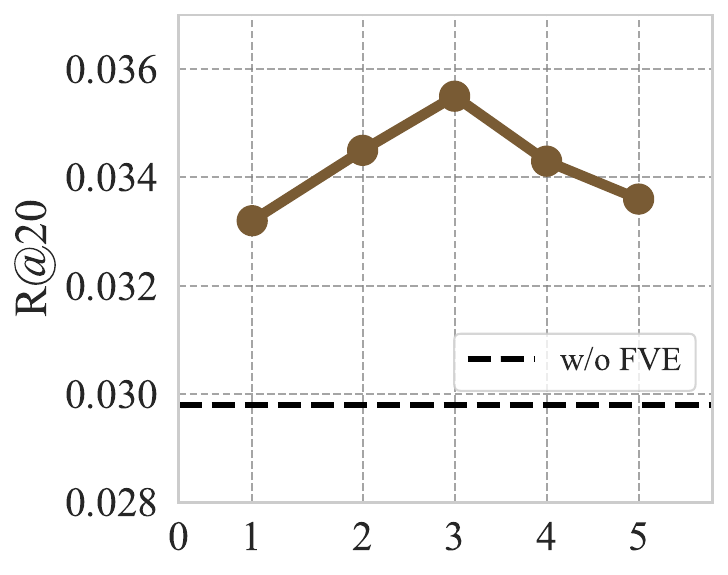}
        \caption{Home Dataset}
    \end{subfigure}
    \hfill
    \begin{subfigure}[t]{0.24\textwidth}
        \centering
        \includegraphics[width=\linewidth]{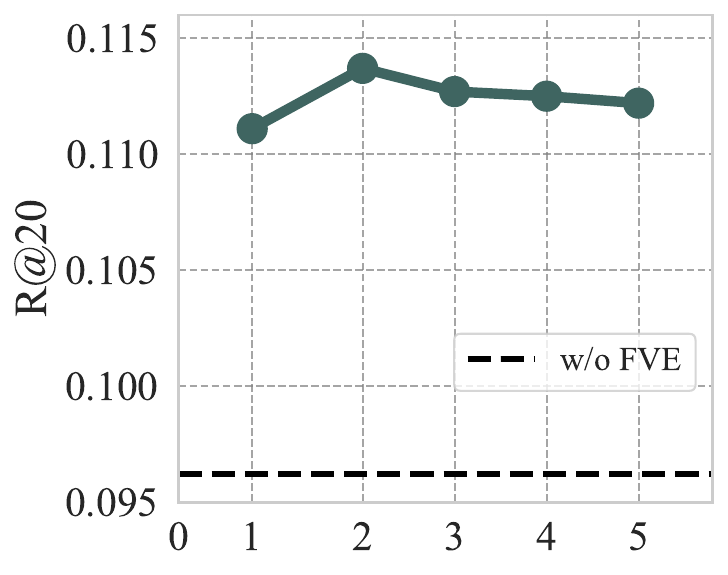}
        \caption{Beauty Dataset}
    \end{subfigure}
    \hfill
    \begin{subfigure}[t]{0.24\textwidth}
        \centering
        \includegraphics[width=\linewidth]{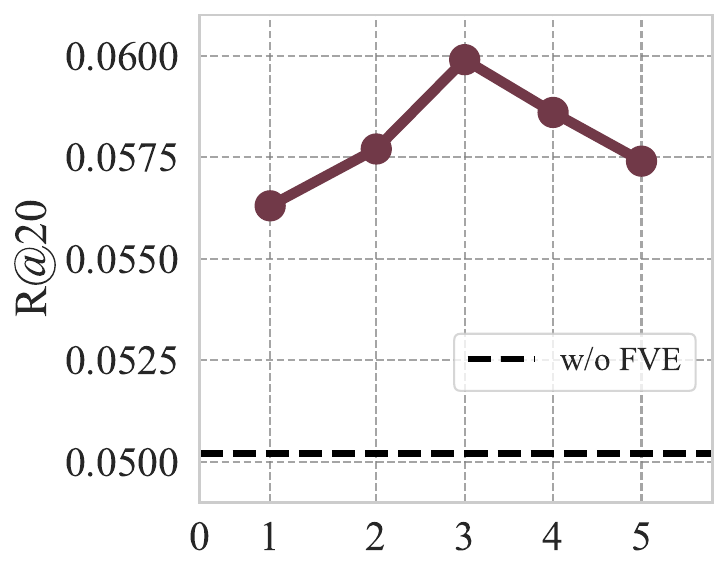}
        \caption{Sports Dataset}
    \end{subfigure}
    \hfill
    \begin{subfigure}[t]{0.24\textwidth}
        \centering
        \includegraphics[width=\linewidth]{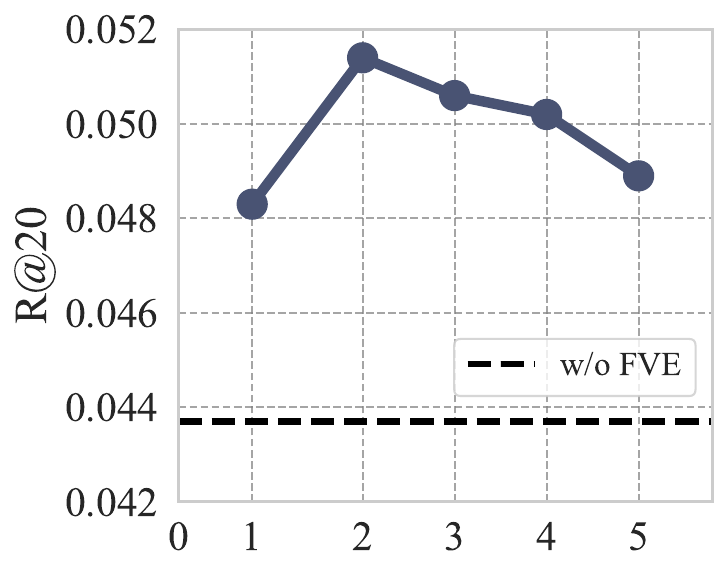}
        \caption{Yelp Dataset}
    \end{subfigure}
    \caption{Performance comparison of M3SRec+REVEAL with $X_{\max}$ hyperparameter settings on the different datasets.}
    \Description{A figure showing hyperparameter $X_{\max}$ comparison on Home, Beauty, Sports, and Yelp datasets.}
    \label{fig:hyper_Xmax}
\end{figure*}

To examine how iterative prompt refinement affects FVE, we vary the refinement cycle number $X_{\max}$ under the Vision-Only setting, where text embeddings are replaced with zero vectors. This setting is adopted to isolate the contribution of FVE more clearly, since the model then relies mainly on ID and visual signals. Here, $X_{\max}$ controls how many times the instantiated prompt is updated using recommendation feedback. Figure~\ref{fig:hyper_Xmax} leads to the following observations.

(1) The effect of $X_{\max}$ is consistently non-monotonic across all datasets. Introducing FVE already yields clear gains over the baseline at $X_{\max}=1$, indicating that even a single round of feedback-guided refinement can improve visual extraction. This suggests that the initial prompt, although coarse, can already be corrected toward more recommendation-relevant visual attributes after one round of feedback. As $X_{\max}$ further increases, performance first improves and then declines, indicating that a moderate number of refinement cycles is beneficial, whereas excessive refinement may introduce unstable or overly specific prompt updates.

(2) The optimal refinement depth is dataset-dependent. Home and Sports achieve their best Recall@20 at $X_{\max}=3$, improving from 0.0332 to 0.0355 and from 0.0563 to 0.0599, respectively. One possible explanation is that these two datasets may involve richer and more diverse visual attributes, such as style, structure, material, or functional design, which are not fully captured by the initial prompt focusing on overall appearance. In addition, their associated reviews may provide more useful attribute-level cues, allowing later refinement cycles to gradually emphasize preference-relevant details rather than only coarse-grained visual impressions. In contrast, Beauty and Yelp peak earlier at $X_{\max}=2$, achieving the best Recall@20 of 0.1137 and 0.0514, respectively. This suggests that the major useful visual signals in these datasets may be captured with fewer refinement cycles. For Beauty, the useful cues may be more concentrated in localized attributes, such as package style, color, or texture-related details, so one or two rounds of refinement may already be sufficient to move beyond overall appearance. For Yelp, the visual modality is likely less dominant than in product-oriented datasets, since user preference may also depend heavily on non-visual factors such as service, taste, or atmosphere, making earlier saturation more likely.

(3) Compared with the Vision-Only baseline, the gains at the best setting remain substantial on all datasets, further demonstrating the importance of high-quality visual extraction when textual information is unavailable. Overall, the results suggest that two to three refinement cycles are sufficient in most cases. In our experiments, the optimal value is $X_{\max}=3$ for Home and Sports, and $X_{\max}=2$ for Beauty and Yelp, indicating that the suitable refinement depth is dataset-dependent.

\subsubsection{Effect of the Calibration Strength $\alpha$}

\begin{figure*}[t]
    \centering
    \begin{subfigure}[t]{0.24\textwidth}
        \centering
        \includegraphics[width=\linewidth]{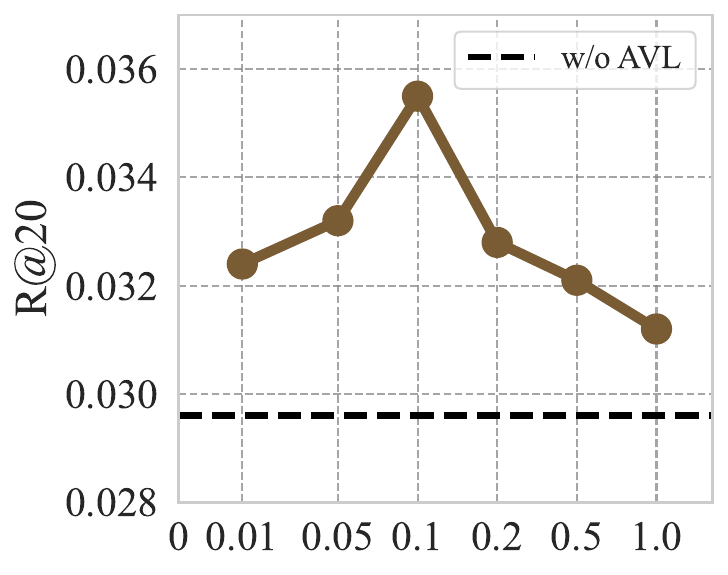}
        \caption{Home Dataset}
    \end{subfigure}
    \hfill
    \begin{subfigure}[t]{0.24\textwidth}
        \centering
        \includegraphics[width=\linewidth]{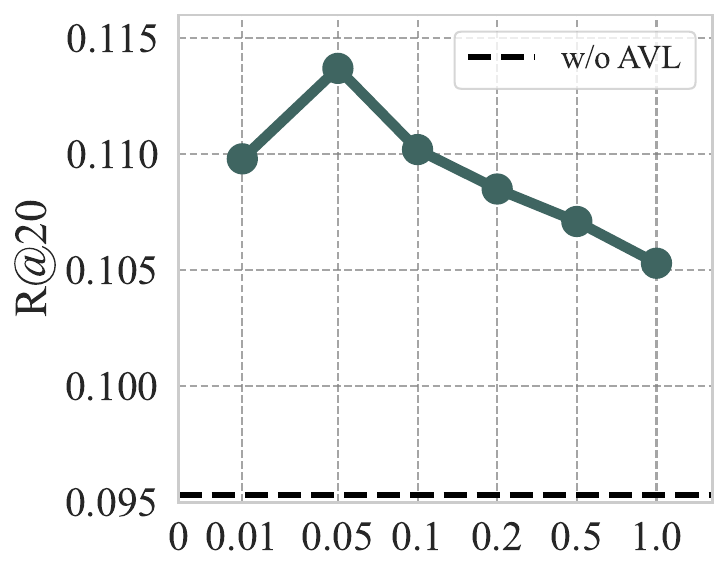}
        \caption{Beauty Dataset}
    \end{subfigure}
    \hfill
    \begin{subfigure}[t]{0.24\textwidth}
        \centering
        \includegraphics[width=\linewidth]{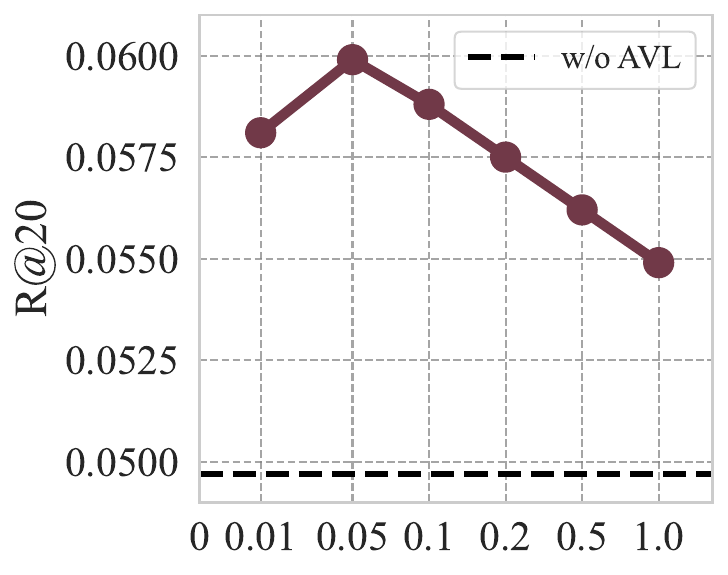}
        \caption{Sports Dataset}
    \end{subfigure}
    \hfill
    \begin{subfigure}[t]{0.24\textwidth}
        \centering
        \includegraphics[width=\linewidth]{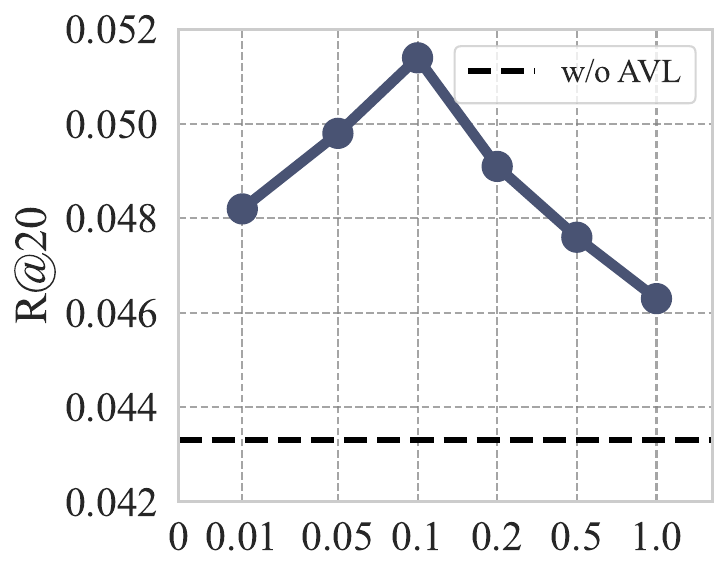}
        \caption{Yelp Dataset}
    \end{subfigure}
    \caption{Performance comparison of M3SRec+REVEAL with $\alpha$ hyperparameter settings on the different datasets.}
    \Description{A figure showing hyperparameter $\alpha$ comparison on Home, Beauty, Sports, and Yelp datasets.}
    \label{fig:hyper_alpha}
\end{figure*}

To examine how the calibration strength affects AVL, we vary $\alpha$ under the Vision-Only setting, where text embeddings are replaced with zero vectors. This setting is adopted to isolate the contribution of AVL more clearly, since the model then relies mainly on ID and visual signals. In this case, adjusting $\alpha$ directly controls how strongly AVL calibrates the gradients of vision-related parameters according to the discrepancy between visual and textual unimodal performance. Figure~\ref{fig:hyper_alpha} leads to the following observations.

(1) The effect of $\alpha$ is consistently non-monotonic across all datasets. Introducing AVL already improves performance over the Vision-Only baseline, which indicates that explicit gradient calibration is beneficial when textual information is unavailable. However, when $\alpha$ is too small, the improvement remains limited, whereas excessively large values lead to clear performance decline. This suggests that AVL requires a moderate calibration strength: insufficient calibration cannot effectively compensate for visual under-optimization, while overly aggressive calibration likely amplifies batch-level noise and weakens training stability.

(2) The optimal calibration strength is dataset-dependent. Home and Yelp achieve their best Recall@20 at $\alpha=0.1$, reaching 0.0355 and 0.0514, respectively. This suggests that these two datasets may require slightly stronger visual gradient calibration, potentially because their visual signals are useful but not sufficiently dominant to be fully exploited without additional optimization emphasis. In contrast, Beauty and Sports peak earlier at $\alpha=0.05$, achieving the best Recall@20 of 0.1137 and 0.0599, respectively. This indicates that for these datasets, a moderate calibration strength is already sufficient, and further increasing $\alpha$ may over-amplify visual updates. One possible explanation is that the useful visual cues in Beauty and Sports may already be relatively clear once a small amount of calibration is introduced, so stronger scaling mainly adds optimization instability rather than additional useful supervision.

(3) Performance drops consistently once $\alpha$ exceeds 0.1, which indicates that the gradient calibration in AVL should remain conservative. For example, Recall@20 on Beauty decreases from 0.1137 at $\alpha=0.05$ to 0.1053 at $\alpha=1.0$, while Sports declines from 0.0599 to 0.0549 over the same range. Similar trends can also be observed on Home and Yelp. Since AVL rescales visual gradients according to the deviation between instantaneous and historical modality gaps, an excessively large $\alpha$ likely makes the update factor too sensitive to local fluctuations, thereby weakening robust optimization. Overall, the results suggest that relatively small values of $\alpha$ are sufficient in most cases. In our experiments, the optimal value is $\alpha=0.1$ for Home and Yelp, and $\alpha=0.05$ for Beauty and Sports, indicating that the suitable calibration strength is dataset-dependent.

\subsection{Computational Cost Analysis (RQ6)}

\begin{table*}[t]
\centering
\caption{The training and inference cost comparison of M3SRec and M3SRec+REVEAL across different datasets. Under \textbf{Training Time (h)}, \textbf{w/o FVE} denotes the backbone training time without the FVE module, \textbf{w/ FVE} denotes the additional training overhead introduced by FVE, and \textbf{Total} denotes the overall training time. M3SRec+REVEAL uses Qwen2.5-VL-7B in FVE. Inference time is reported in seconds under the same deployment setting.}
\label{tab:cost_analysis}
\resizebox{0.8\linewidth}{!}{
\begin{tabular}{c|c|c|c|c|c|c}
\toprule
\multirow{2}{*}{\textbf{Dataset}} 
& \multirow{2}{*}{\textbf{Model}} 
& \multicolumn{3}{c|}{\textbf{Training Time (h)}} 
& \multirow{2}{*}{\makecell{\textbf{Inference} \\ \textbf{Time (s)}}}
& \multirow{2}{*}{\textbf{Memory (GB)}} \\
\cline{3-5}
&  & \textbf{w/o FVE} & \textbf{w/ FVE} & \textbf{Total} &  &  \\
\midrule
\multirow{2}{*}{Home}
& M3SRec & 0.45 & -- & 0.45 & 15.2 & 2.75 \\
\cline{2-7}
& M3SRec+REVEAL & 0.52 & 8.10 & 8.62 & 15.2 & 2.86 \\
\midrule
\multirow{2}{*}{Beauty}
& M3SRec & 0.16 & -- & 0.16 & 1.3 & 2.60 \\
\cline{2-7}
& M3SRec+REVEAL & 0.18 & 2.20 & 2.38 & 1.3 & 2.92 \\
\midrule
\multirow{2}{*}{Sports}
& M3SRec & 0.23 & -- & 0.23 & 4.1 & 2.66 \\
\cline{2-7}
& M3SRec+REVEAL & 0.27 & 5.57 & 5.84 & 4.1 & 2.84 \\
\midrule
\multirow{2}{*}{Yelp}
& M3SRec & 4.65 & -- & 4.65 & 152.6 & 7.51 \\
\cline{2-7}
& M3SRec+REVEAL & 5.20 & 17.96 & 23.16 & 152.6 & 8.01 \\
\bottomrule
\end{tabular}
}
\end{table*}

Table~\ref{tab:cost_analysis} reports the practical training and inference cost of REVEAL on M3SRec across different datasets. Since REVEAL is applied only during training, its additional cost mainly arises from the FVE refinement process and the extra gradient calibration introduced by AVL. The results lead to the following observations.

(1) REVEAL mainly increases training cost, while leaving inference time nearly unchanged. More importantly, the additional training overhead is dominated by FVE rather than the backbone training itself. Across all datasets, the \textit{w/o} FVE time of M3SRec+REVEAL remains close to that of the original M3SRec backbone. For example, on Home, the backbone-related training time increases only slightly from 0.45 h to 0.52 h, whereas the additional \textit{w/} FVE overhead reaches 8.10 h. A similar pattern can also be observed on Sports, where the backbone-related time is 0.27 h but the FVE overhead is 5.57 h. This result indicates that the main computational burden of REVEAL is not caused by the recommendation backbone itself, but by the repeated VLM-based prompt refinement procedure in FVE. In contrast, inference time remains unchanged across all datasets, which suggests that REVEAL introduces training-stage overhead rather than deployment-stage latency.

(2) The absolute computational cost remains strongly affected by dataset scale, but the proportion of FVE overhead is consistently dominant. Yelp is the most expensive setting, with total training time increasing from 4.65 h to 23.16 h, among which 17.96 h is introduced by FVE. One likely reason is that Yelp contains substantially more users, items, and interactions than the other datasets, which already makes the backbone training itself much more expensive. Consequently, once FVE performs refinement on top of this larger-scale setting, the absolute overhead becomes even more pronounced. In contrast, Beauty remains the lightest setting, where the total training time of M3SRec+REVEAL is only 2.38 h. Nevertheless, even on Beauty, the majority of the additional cost still comes from FVE rather than backbone optimization. These results suggest that the practical cost of REVEAL is jointly determined by dataset scale and the iterative prompt-refinement design, while the latter is the more direct source of overhead.

(3) From a practical perspective, REVEAL introduces a clear training-stage cost--effectiveness trade-off. On the one hand, the substantial FVE overhead indicates that improving visual extraction through repeated VLM-guided prompt refinement is computationally expensive, especially on larger datasets. On the other hand, the unchanged inference time and the relatively small increase in memory usage indicate that this cost is concentrated in offline training and does not affect online deployment efficiency. This distinction is important because it suggests that REVEAL may be particularly suitable for scenarios where recommendation quality is prioritized and additional offline optimization cost is acceptable. More broadly, the results indicate that future work on REVEAL should likely focus on reducing the cost of FVE, for example by decreasing refinement frequency or adopting more lightweight prompt-update strategies, since this component is the primary bottleneck in practice.

\section{Discussion and Conclusion}

This work examines why visual modality often contributes only marginally in multimodal sequential recommendation. The results suggest that this limitation stems not only from noisy visual content, but also from two deeper factors: pretrained visual encoders often fail to capture preference-relevant cues, and textual signals tend to dominate optimization during training. Our study addresses these issues through two complementary mechanisms, namely feedback-guided visual extraction and adaptive visual learning, and extensive experiments show that this design consistently improves multiple MSR backbones while preserving the original inference pipeline.

These findings have several implications. First, the limited contribution of visual modality in current MSR models should not be taken to mean that visual information is inherently weak. Rather, it likely indicates that existing models do not extract or optimize visual signals in a recommendation-aware manner. Second, the results suggest that frozen generative VLMs can serve as controllable visual extractors rather than static feature providers, since recommendation feedback can redirect visual focus through prompt refinement without requiring parameter updates. Third, the gains brought by adaptive optimization indicate that modality imbalance is not only an architectural issue, but also an optimization issue: even useful visual signals may remain underutilized unless their learning process is explicitly strengthened.

Several directions remain for future work. A first limitation is training cost, since iterative prompt refinement and unimodal calibration introduce non-negligible overhead. More efficient variants, such as reducing refinement frequency or distilling refined visual guidance into lightweight encoders, would therefore be valuable. A second limitation is that the current refinement process is still relatively coarse-grained, as prompts are updated only after each complete backbone training stage. Finer-grained yet stable refinement strategies may further improve adaptability. In addition, our results suggest that larger VLMs do not always yield consistent gains, indicating that recommendation-oriented visual alignment may matter more than model scale alone. Finally, although this work focuses on visual modality, the same idea of feedback-guided extraction and modality-aware optimization may potentially be extended to other modalities or richer multimodal recommendation settings.

In conclusion, the main message of this paper is that stronger multimodal recommendation does not simply depend on introducing more modalities, but on enabling each modality to contribute in a preference-aware and optimization-aware manner. By improving both visual extraction and visual learning, the proposed REVEAL framework offers a practical way to better exploit visual information in MSR and provides a useful direction for future multimodal recommendation research.

\bibliographystyle{ACM-Reference-Format}
\bibliography{TOIS}

\end{document}